\documentclass{sig-alternate-10pt}
\pdfoutput=1
\usepackage{times}
\usepackage{epsfig,endnotes}
\usepackage{url}
\usepackage{xspace} 
\usepackage{amsmath}
\usepackage{mathtools}
\usepackage{subfigure}
\usepackage{empheq}
\usepackage[table]{xcolor}
\usepackage{verbatim}

\usepackage{booktabs}
\usepackage{tabularx}
\usepackage{pifont}
\usepackage{multirow}
\newcommand{\ra}[1]{\renewcommand{\arraystretch}{#1}}

\begin{document}

\title{\Large \bf Performance Characterization of a Commercial Video Streaming Service}

\numberofauthors{5}
\author{
\alignauthor 
Mojgan Ghasemi\\
\affaddr{Princeton University}
\alignauthor 
Partha Kanuparthy\\
\affaddr{Yahoo Research}
\alignauthor 
Ahmed Mansy\\
\affaddr{Yahoo}
\and
Theophilus Benson\\
\affaddr{Duke University}
\alignauthor
Jennifer Rexford\\
\affaddr{Princeton University}
}

\maketitle 
\thispagestyle{empty}

\subsection*{Abstract}
Despite the growing popularity of video streaming over the Internet, problems such as re-buffering and high startup latency continue to plague users.  In this paper, we present an end-to-end characterization of Yahoo's video streaming service, analyzing over 500 million video chunks downloaded over a two-week period.  We gain unique visibility into the causes of performance degradation by instrumenting both the CDN server and the client player at the chunk level, while also collecting frequent snapshots of TCP variables from the server network stack.  We uncover a range of performance issues, including an asynchronous disk-read timer and cache misses at the server, high latency and latency variability in the network, and buffering delays and dropped frames at the client.  Looking across chunks in the same session, or destined to the same IP prefix, we see how some performance problems are relatively persistent, depending on the video's popularity, the distance between the client and server, and the client's operating system, browser, and Flash runtime.
\section{Introduction}
Internet users watch hundreds of millions of videos per day~\cite{youtubeStats}, and video streams represent more than 70\% of North America's downstream traffic during peak hours~\cite{Sandvine}. A video streaming session, however, may suffer from problems such as long startup delay, re-buffering events, and low  video quality that negatively impact user experience and decrease the content provider's revenue~\cite{Krishnan:2012:VSQ:2398776.2398799,Dobrian:2011:UIV:2043164.2018478}. 
Content providers strive to improve performance through a variety of optimizations, such as placing servers ``closer'' to clients, content caching, effective peering and routing decisions, and splitting a session into fixed-length \emph{chunks} available in a variety of bitrates to allow adaptive bitrate algorithms (ABR) in the player to adjust video quality to available resources~\cite{Akhshabi:2011:EER:1943552.1943574,Yin:2015:CAD:2785956.2787486,Huang:2014:BAR:2619239.2626296,Jiang:2012:IFE:2413176.2413189,Tian:2012:TAS:2413176.2413190}.

\begin{figure}
  \includegraphics[width =0.85\linewidth]{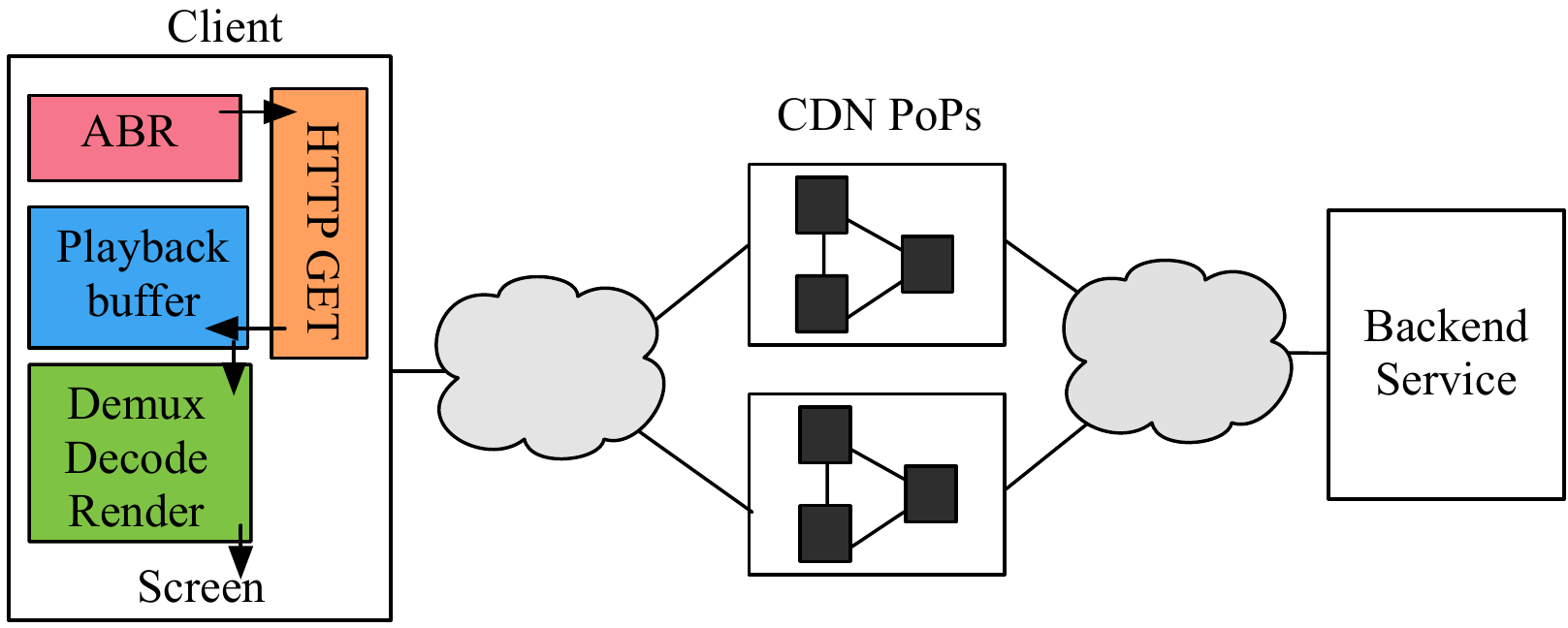}
  \caption{End-to-End video delivery components.}
  \label{f:components}
\end{figure}

Despite these optimizations, performance problems can arise anywhere along
the end-to-end delivery path shown in Figure~\ref{f:components}. The bad performance
can stem from a variety of root causes. For example, the \emph{backend service} may increase the chunk download latency on a cache miss. The \emph{CDN} servers can introduce high latency in accessing data from disk. The \emph{network} can introduce congestion or random packet losses. The \emph{client's download stack} may handle data inefficiently (e.g., slow copying of data from OS to the player via the browser and Flash runtime) and the \emph{client's rendering path} may drop some frames due to high CPU load.

While ABRs can adapt to performance problems (e.g., lower the bitrate when throughput is low), understanding the \emph{location} and \emph{root causes} of performance problems enables content providers to take the right corrective (or even proactive) actions, such as directing client requests to different servers, adopting a different cache-replacement algorithm, or further optimizing the player software.  In some cases, knowing the bottleneck can help the content provider decide \emph{not} to act, because the root cause is beyond the provider's control---for example, it lies in the client's browser, operating system, or access link.  The content provider could detect the existence of performance problems by collecting Quality of Experience (QoE) metrics at the player, but this does not go far enough to identify the underlying cause. In addition, the buffer at the player can (temporarily) mask underlying performance problems, leading to delays in detecting significant problems based solely on QoE metrics.

Instead, we adopt a \emph{performance-driven} approach for uncovering performance problems.  Collecting data at the client or the CDN alone is not enough.  Client-side measurements, while crucial for uncovering problems in the download stack (e.g., a slow browser) or rendering path (e.g., slow decoder), cannot isolate network and provider-side bottlenecks. Moreover, a content provider cannot collect OS-level logs or measure the network stack at the client; even adding small extensions to the browsers or plugins would complicate deployment.  Server-side logging can fill in the gaps~\cite{Yu:2011:PNP:1972457.1972464}, with care to ensure that the measurements are sufficiently lightweight.

In this paper, we instrument the CDN servers and video player of a Web-scale commercial video streaming service, and join the measurement data to construct an end-to-end view of session performance.  We measure per-\emph{chunk} milestones at the player which runs on top of Flash (e.g., the time to get the chunk's first and last bytes, and the number of dropped frames during rendering), and the CDN server (e.g., server and backend latency), as well as kernel-space TCP variables (e.g., congestion window and round-trip time) from the server host. Direct observation of the main system components help us avoid relying on inference or tomography techniques that would limit the accuracy, or relying on some other source of ``ground truth'' to label the data for machine learning~\cite{Dim2015}.
In this work we make the following contributions: 

1. A large-scale instrumentation of \emph{both sides} of the video delivery path in a commercial video streaming service over a two-week period, studying more than 523 million chunks and 65 million on-demand video sessions. 
 
2. End-to-end instrumentation that allows us to characterize the player, network path and the CDN components of session performance, across multiple layers of the stack, \emph{per-chunk}. 
We show an example of how partial instrumentation (e.g., player-side alone) would lead to incorrect conclusions about performance problems. Such conclusions could cause the ABR algorithm to make wrong decisions.

3. We characterize transient and persistent problems in the end-to-end path that have not been studied before; in particular the client's \emph{download stack} and \emph{rendering path}, and show their impact on QoE.

4. We offer a \emph{comprehensive characterization} of performance problems for Internet video, and our key findings are listed in Table~\ref{t:findings}. Based on these findings, we offer insights for video content providers and Internet providers to improve video QoE.

\begin{table}\centering
\small
\begin{tabularx}{\linewidth}{p{1cm}p{7cm}}
\textbf{Location}&\textbf{Findings}\\ \toprule
\textbf{CDN} & 1. Asynchronous disk reads increases server-side delay.\\ 
& 2. Cache misses increase CDN latency by order of magnitude.\\ 
& 3. Persistent cache-miss and slow reads for unpopular videos.\\ 
& 4. Higher server latency even on lightly loaded machines.\\
\hline
\textbf{Network} & 1. Persistent delay due to physical distance or enterprise paths.\\
%
%
%
& 2. Higher latency variation for users in enterprise networks.\\
& 3. Packet losses early in a session have a bigger impact.\\
& 4. Bad performance caused more by throughput than latency.\\
\hline
\textbf{Client} & 1. Buffering in client download stack can cause re-buffering. \\
& 2. First chunk of a session has higher download stack latency.\\
& 3. Less popular browsers drop more frames while rendering.\\ 
& 4. Avoiding frame drops need min of $1.5\frac{sec}{sec}$ download rate.\\
& 5. Videos at lower bitrates have \emph{more} dropped frames.\\
\hline
\bottomrule
\end{tabularx}
\caption{Summary of key findings.}
\label{t:findings}
\end{table}

\section{Chunk Performance Monitoring}
\label{s:chunk}

\noindent\textbf{Model.} Our model of a video session is a linearizable sequence of HTTP(S)~\footnote{Both HTTP and HTTPS protocols are supported, for simplicity we will only use HTTP in the rest of the paper} requests and responses over the same TCP connection between the player and the CDN server, after the player has been assigned to a server. The session starts with the player requesting the manifest, which contains a list of chunks in available bitrates (upon errors and user events such as seeks manifest is requested again). The ABR algorithm, that has been tuned and tested in the wild to balance between low startup delay, low re-buffering rate, high quality and smoothness, chooses a bitrate for each chunk to be requested from the CDN server. The CDN service maintains a FIFO queue of arrived requests and maintains a threadpool to serve the queue. The CDN uses a ``multi-level'' and distributed cache (between the main memory and the local disk) to cache chunks with an LRU replacement policy, and upon a cache miss, makes a corresponding request to the backend service. 

The client host includes two independent execution paths that share host resources. The \emph{download} path ``moves'' chunks from the NIC to the player, by writing them to the playback buffer. The \emph{rendering} path reads from the playback buffer, de-muxes (audio from video), and decodes and renders the pixels on the screen---this path could use either the GPU or the CPU. Note that there is a stack below the player: the player executes on top of a Javascript and Flash runtime, which in turn is run by the browser on top of the OS.

\subsection{Chunk Instrumentation}

We collect \emph{chunk}-level measurements because: (1) most decisions affecting performance are taken per-chunk (e.g., caching at the CDN, and bitrate selection at the player), although some metrics are chosen once per session (e.g., the CDN server),  (2) sub-chunk measurements would increase CPU load on client, at the expense of rendering performance (Section~\ref{s:render}), and (3) client-side handling of data within a chunk can vary across streaming technologies, and is often neither visible nor controllable. For example, players implemented on top of Flash use a progress event that delivers data to the player, and 
the buffersize or frequency of this event may vary across browsers or versions.

\begin{figure}
\centering
  \includegraphics[width=0.8\linewidth]{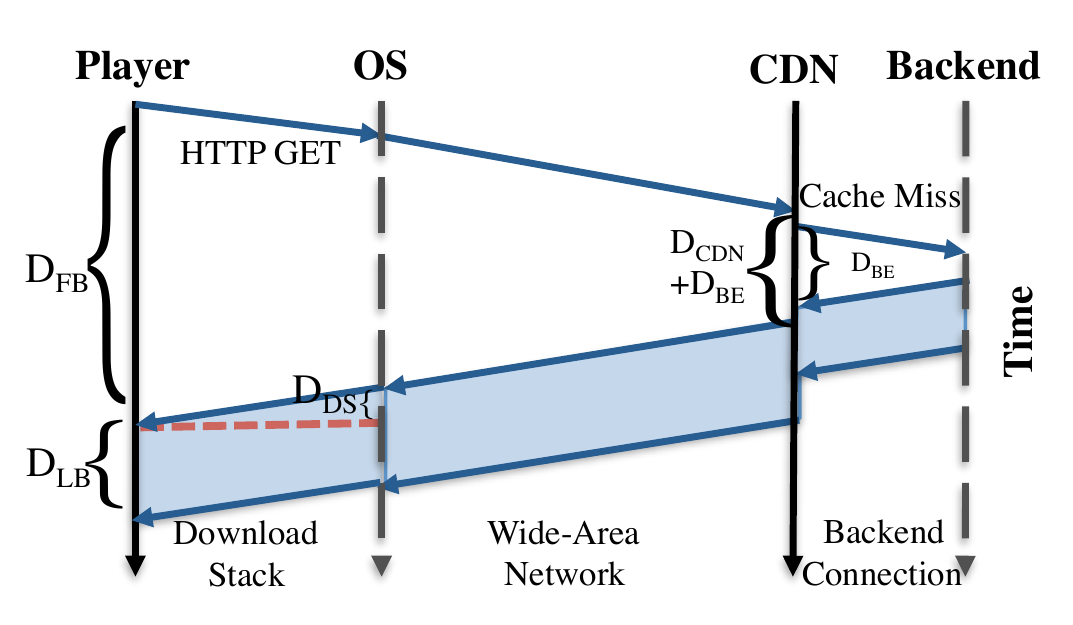}
  \caption{Time diagram of chunk delivery. Solid lines are instrumentation while dashed lines are estimates. }
  \label{f:timediagram}
\end{figure}

We capture the following milestones per-chunk at the player and the CDN service: (1) When the chunk's HTTP GET request is sent, (2) CDN latency in serving the chunk, in addition to backend latency for cache misses, and (3) the time to download the first and last bytes of the chunk. We denote the player-side \emph{first-byte delay} $D_{FB}$ and \emph{last-byte delay} $D_{LB}$. Figure~\ref{f:timediagram} summarizes our notation.
We divide a chunk's lifetime into the three phases: fetch, download, and playout.

\vspace{0.1in}\noindent\textbf{Fetch phase.} The fetch process starts with the player sending an HTTP request to the CDN for a chunk at a specified bitrate until the first byte arrives at the player. The byte transmission and delivery traverses the host stack (player, Flash runtime, browser, userspace to kernel space and the NIC)---contributing to the \emph{download stack latency}. If the content is cached at the CDN server, the first byte is sent after a delay of $D_{CDN}$ (the cache lookup and load delay); otherwise, the backend request for that chunk incurs an additional delay of $D_{BE}$. Note that the backend and delivery are always pipelined. The first-byte delay $D_{FB}$ includes network round-trip time ($rtt_0$), CDN service latency, backend latency (if any), and client download stack latency:

\begin{equation}
D_{FB} = D_{CDN} + D_{BE} + D_{DS} + rtt_0
\label{e:rtt0}
\end{equation}

We measure $D_{FB}$ for each chunk at the player. At the CDN service, we measure $D_{CDN}$ and its constituent parts: 
(1) $D_{wait}$: the time the HTTP request waits in the queue until the request headers are read by the server,
(2) $D_{open}$: after the request headers are read until the server first attempts to open the file, regardless of cache status, and
(3) $D_{read}$: time to read the chunk's first byte and write it to the socket, including the delay to read from local disk or backend.
The backend latency ($D_{BE}$) is measured at the CDN service and includes network delay. Characterizing backend service problems is out of scope for this work; such problems are relatively rare.

A key limitation of player-side instrumentation is that it captures the mix of download stack latency, network latency, and server-side latency. To isolate network performance from end-host performance, we measure the end-to-end network path at the CDN host kernel's TCP stack. Since kernel-space latencies are relatively very low, it is reasonable to consider this view as representative of the network path performance. Specifically, the CDN service snapshots the Linux kernel's \texttt{tcp\_info} structure for the player TCP connection (along with context of the chunk being served). The structure includes TCP state such as smoothed RTT, RTT variability, retransmission counts, and sender congestion window. We sample the path performance periodically every 500ms\footnote{The frequency is chosen to keep overhead low in production.}; this allows us to observe changes in path performance.


\vspace{0.1in}\noindent\textbf{Download Phase.}
The download phase is the window between arrivals of the first and the last bytes of the chunk at the player, i.e., the last-byte delay. $D_{LB}$ depends on the chunk size, which depends on chunk bitrate and duration. To identify chunks suffering from low throughput, on the client side we record the requested \emph{bitrate} and the \emph{last-byte delay}. To understand the network path performance and its impact on TCP, we snapshot TCP variables from the CDN host kernel at least once per-chunk (as described above).

\vspace{0.1in}\noindent\textbf{Playout phase.}
As a chunk is downloaded, it is added to the playback buffer. If the playback buffer does not contain enough data, the player pauses and waits for sufficient data; in case of an already playing video, this causes a rebuffering event. We instrument the player to measure the number ($buf_{count}$) and duration of rebuffering events ($buf_{dur}$) per-chunk played.

Each chunk must be decoded and rendered at the client. In the absence of hardware rendering (i.e., GPU), chunk frames are decoded and rendered by the CPU, which makes video quality sensitive to CPU utilization. A slow rendering process drops frames to keep up with the encoded frame rate. To characterize rendering path problems, we instrument the player to collect the average rendered frame rate per chunk ($avg_{fr}$) and the number of dropped frames per chunk ($drop_{fr}$). A low rendering rate, however, is not always indicative of bad performance; for example, when the  player is in a hidden tab or a minimized window, video frames are dropped to reduce CPU load \cite{Dobrian:2011:UIV:2043164.2018478}. To identify these scenarios, the player collects a variable ($vis$) that records if the player is visible when the chunk is displayed.  Table~\ref{t:instrumentations} summarizes the metrics collected for each chunk at the player and CDN.

\begin{table}\centering
\small
\ra{1.3}
\begin{tabularx}{\linewidth}{p{2.2cm}X}
\textbf{Location} & \textbf{Statistics} \\ \toprule
Player (Delivery) & sessionID, chunkID, $D_{FB}$, $D_{LB}$, bitrate\\
\hline
Player (Rendering) & $buf_{dur}$, $buf_{count}$, $vis$, $avg_{fr}, drop_{fr}$\\
\hline
CDN (App layer) & sessionID, chunkID, $D_{CDN}$ (wait, open, and read), $D_{BE}$, cache status, chunk size\\ 
\hline
CDN (TCP layer) & CWND, SRTT, SRTTVAR, retx, MSS\\
\bottomrule
\end{tabularx}
\caption{Per-chunk instrumentation at player and CDN.}
\label{t:instrumentations}
\end{table}

\subsection{Per-session Instrumentation} 
In addition to per-chunk milestones, we collect session metadata; see Table~\ref{t:session}. A key to end-to-end analysis is to \emph{trace} session performance from the player through the CDN (at the granularity of chunks). We implement tracing by using a globally unique session ID and per-session chunk IDs.

\begin{table}\centering
\small
\ra{1.3}
\begin{tabularx}{\linewidth}{p{2cm}X}
\textbf{Location} & \textbf{Statistics} \\ \toprule
Player & sessionID, user IP, user agent, video length \\
\hline
CDN & sessionID, user IP, user agent, CDN PoP, CDN server, AS, ISP, connection type,  location \\ 
\bottomrule
\end{tabularx}
\caption{Per-session instrumentation at player and CDN.}
\label{t:session}
\end{table}

\section{Measurement Dataset}
We study 65 million VoD sessions (523m chunks) with Yahoo collected over a period of 18 days in September 2015. These sessions were served by a random subset of 85 CDN servers across the US. Our dataset predominantly consists of clients in North America (over 93\%).
 
\begin{figure}[t!]
\centering
\subfigure[CCDF of video lengths (one month)]{\label{f:both:a}\includegraphics[width=0.45\linewidth]{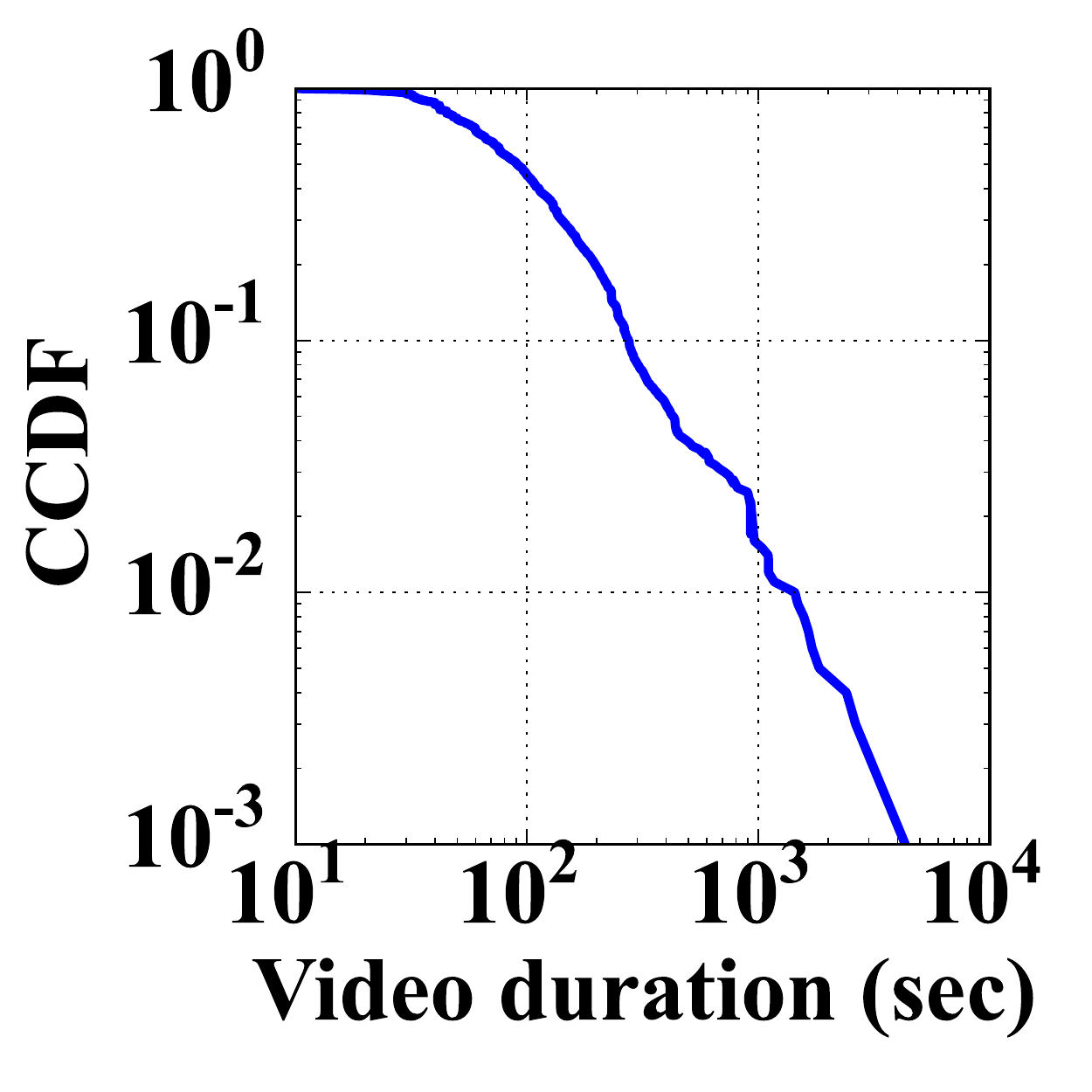}}
\subfigure[Rank vs. popularity (one day)]{\label{f:both:b}\includegraphics[width=0.45\linewidth]{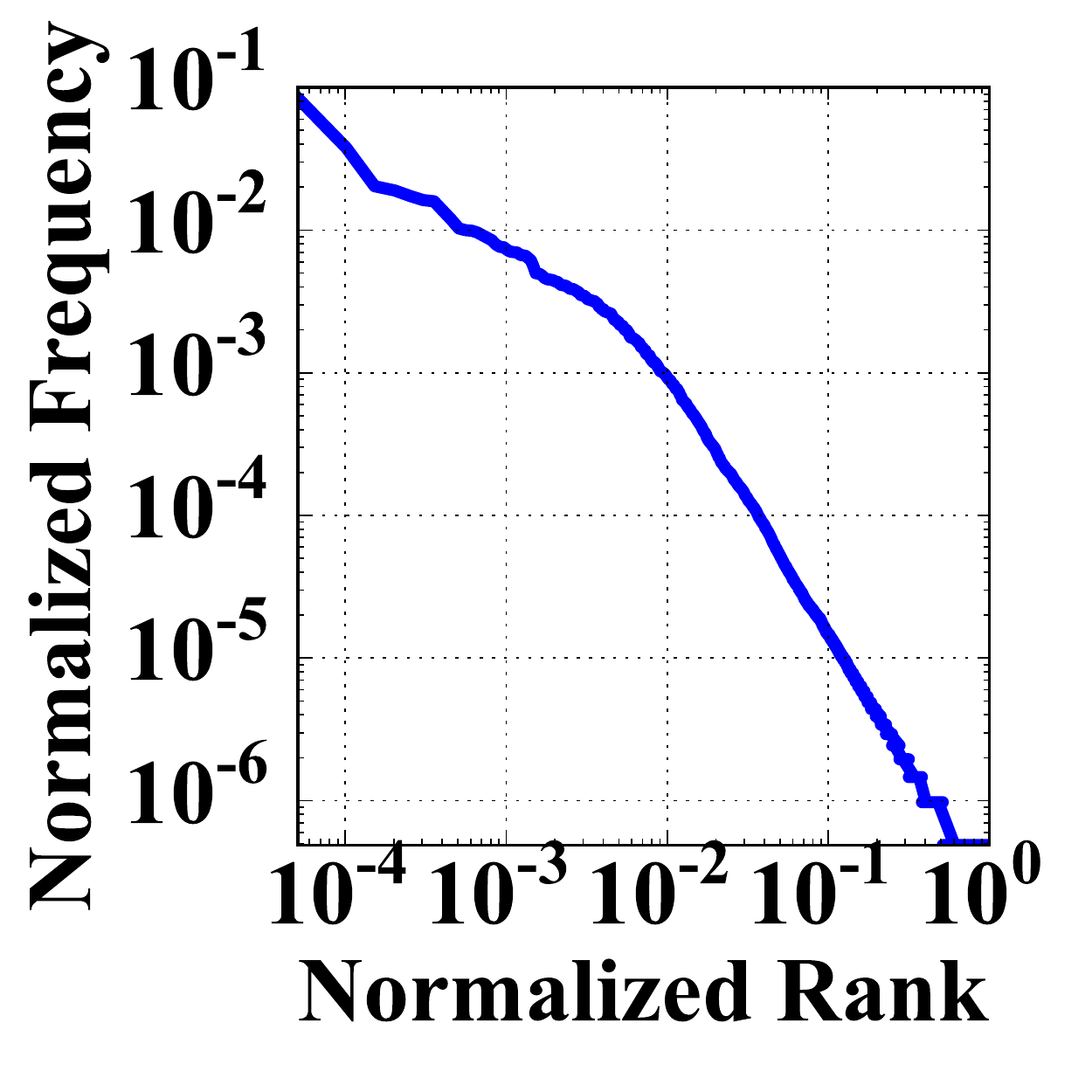}}
\caption{Length and popularity of videos in the dataset.}
\label{f:both}
\end{figure}

Figure~\ref{f:both:a} shows the cumulative distribution of the length of the videos. All chunks in our dataset contain six seconds of video (except, perhaps, the last chunk). 

We focus on desktop and laptop sessions with Flash-based players. The browser distribution is as follows: 43\% Chrome, 37\% Firefox, 13\% Internet Explorer, 6\% Safari, and about 2\% other browsers; the two major OS distributions in the data are Windows (88.5\% of sessions) and OS X (9.38\%). We do not consider cellular users in this paper.



The video viewership and popularity of videos is heavily skewed towards popular content,  
conventional wisdom based on web objects \cite{??}; 
see Figure~\ref{f:both:b}. We find that top 10\% of most popular videos receive about 66\% of all playbacks.

\noindent\textbf{Data preprocessing to filter proxies.}
A possible pitfall in our analysis is the existence of enterprise or ISP HTTP proxies~\cite{Xu15a}, since the CDN server's TCP connection would terminate at the proxy, leading to network measurements (e.g., RTT) reflecting the server-proxy path instead of the client. We filter sessions using a proxy when: (i) we see different
client IP addresses or user agents~\cite{Weaver:2014:HWP:2722265.2722288} between HTTP requests and client-side beacons, or (ii) the client IP address appears in a very large number of sessions (e.g., more more minutes of video per day than there are minutes in a day).
After filtering proxies, our dataset consists of 77\% of sessions.

\noindent\textbf{Ethical considerations:}
Our instrumentation methodology is based on logs/metrics about the traffic, without looking at packet payload or video content. For privacy reasons, we do not track users (through logging) hence we cannot study access patterns of individual users. Our analysis software uses client IP addresses internally to identify proxies and perform coarse-grained geo-location; after that, opaque session IDs are used to join and study the dataset.

\section{Characterizing Performance}
In this section, we characterize the performance of  each component of the end-to-end path, and show the impact on QoE. Prior work~\cite{Dobrian:2011:UIV:2043164.2018478,Yin:2015:CAD:2785956.2787486} has showed that important factors affecting QoE are startup delay, re-buffering ratio, average bitrate, and the rendering quality. 

\subsection{Server-side performance problems}
Yahoo uses the Apache Traffic Server (ATS), a popular caching proxy server~\cite{ats}, to serve HTTP requests.
The traffic engineering system maps clients to CDN nodes using a function of geography, latency, load, cache likelihood, etc. In other words, the system tries to route clients to the server that is likely to have a hot cache. The server first checks the main memory cache, then tries the disk, and finally sends a request to a backend server if needed.

\begin{figure}
\centering
\includegraphics[height=5cm]{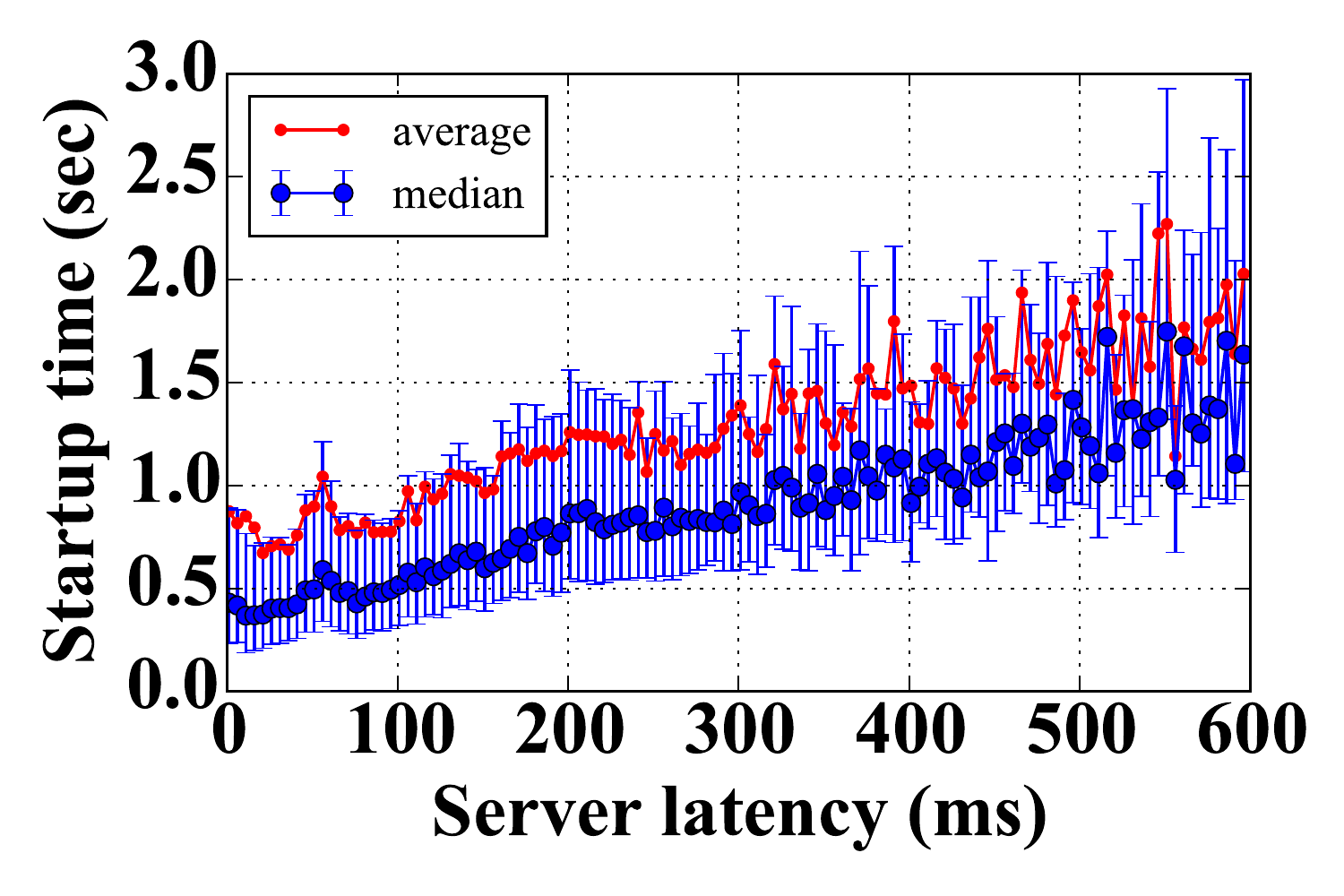}
\caption{Impact of server latency on QoE (startup time), error bars show the interquartile range (IQR).}
\label{f:server_joint}
\end{figure}

Server latencies are relatively low, since the CDN and the backend are well-provisioned. About $5\%$ of sessions, however, experience a QoE problem due to the server, and the problems can be \emph{persistent} as we show below. Figure~\ref{f:server_joint} shows the impact of the server-side latency for the first chunk on the startup delay (time to play) at the player.
 
\begin{figure}[t!]
\centering
\includegraphics[height=5cm]{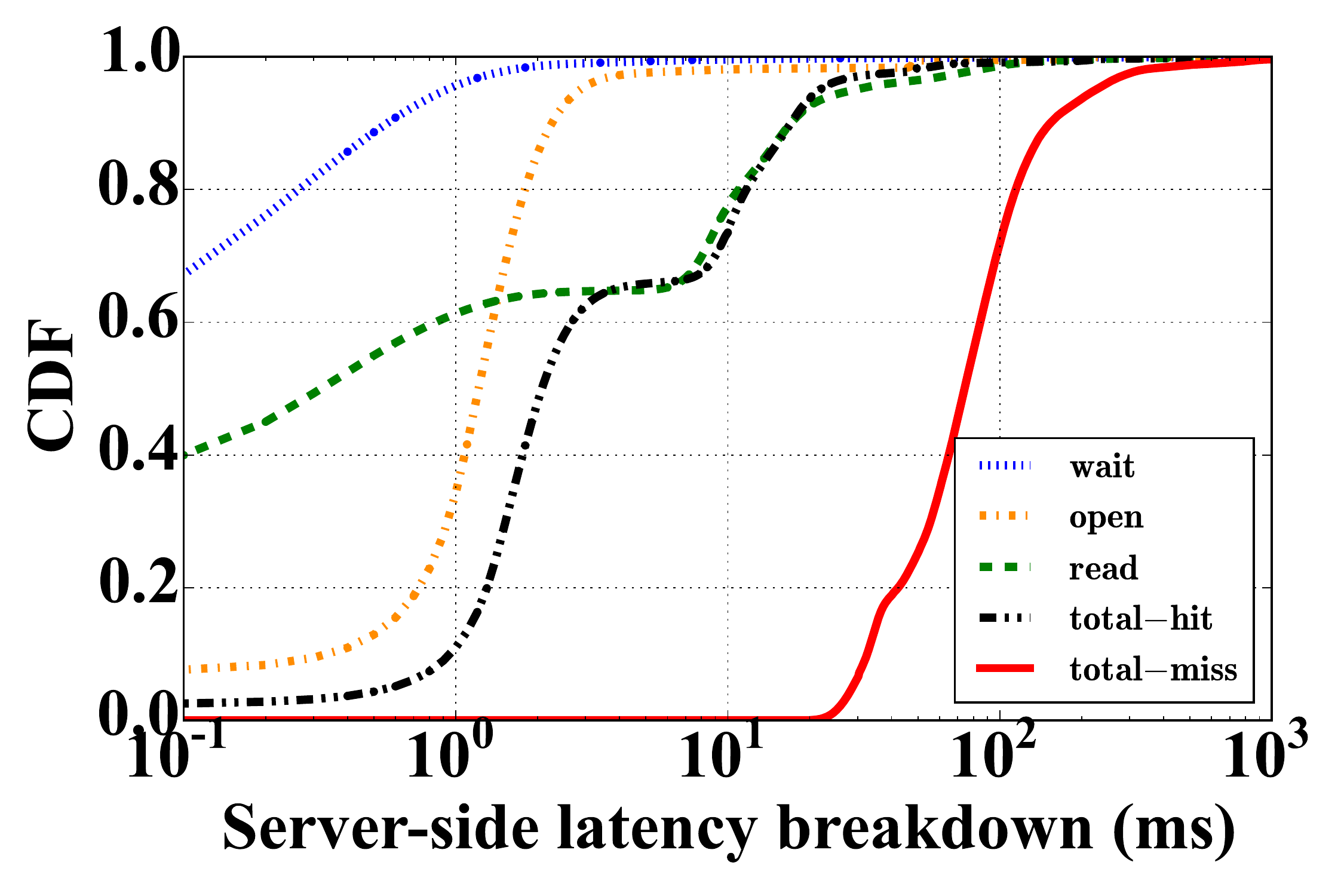}
\caption{CDN latency breakdown across all chunks.}
\label{f:cdn}
\end{figure}

\vspace{0.1in}\noindent\textbf{1. Asynchronous disk read timer and cache misses cause high server latency.}
Figure~\ref{f:cdn} shows the distribution of each component of CDN latency across chunks; it also includes the distribution of total server latency for chunks broken by cache hit and miss. Most of the chunks have a negligible waiting delay ($D_{wait} < 1ms$) and open delay. 
However, the $D_{read}$ distribution has two nearly identical parts, separated by about 10ms. The root cause is that ATS performs an asynchronous read to read the requested files in the background. 
When the first attempt in opening the cache is not immediately returned (due to content not being in memory), retrying to open the file (either from the disk or requesting it from backend service) uses a 10ms~\cite{timer}.

On a cache miss, the backend latency significantly affects the serving latency according to Figure~\ref{f:cdn}. The median server latency among chunks experiencing a cache hit is 2ms, while the median server latency for cache misses is 40 times higher at 80ms. The average and $95^{th}$ percentile of server latency in case of cache misses is ten times more. In addition, cache misses are the main contributor when server latency has a higher contribution to $D_{FB}$ than the network RTT: for 95\% of chunks, network latency is higher than server latency; however, among the remaining 5\%, the cache miss ratio is 40\%, compared to an average cache miss rate of 2\% across session chunks.

\noindent\textbf{Take-away:}
Cache misses impact serving latency, and hence QoE (e.g., startup time) significantly. To offer better cache hit rates, the default LRU cache eviction policy in ATS could be changed to better suited policies for popular-heavy workloads such as GD-size or perfect-LFU~\cite{749260}.

\vspace{0.1in}\noindent\textbf{2. Less popular videos have persistent problems.}
We observed that a small fraction of sessions experience performance problems that are \emph{persistent}.  Once a session has a cache miss on one chunk, the chance of further cache misses increases dramatically; the mean cache miss ratio among sessions with at least one cache miss is $60\%$ (median of $67\%$). Also, once a session has at least one chunk with a high latency ($>10ms$), the chance of future read delays increases; the mean ratio of high-latency chunks in sessions with at least one such chunk is $60\%$ (median of $60\%$). .eps

One possible cause for persistent latency even when the cache hit ratio is high, is a highly loaded server that causes high serving latency
; however, our analysis shows that server latency is not correlated with load.\footnote{We estimated load as of number of parallel HTTP requests, sessions, or bytes served per second.} This is because the CDN servers are well provisioned to handle the load.

\begin{figure}
\centering
\includegraphics[width=\linewidth]{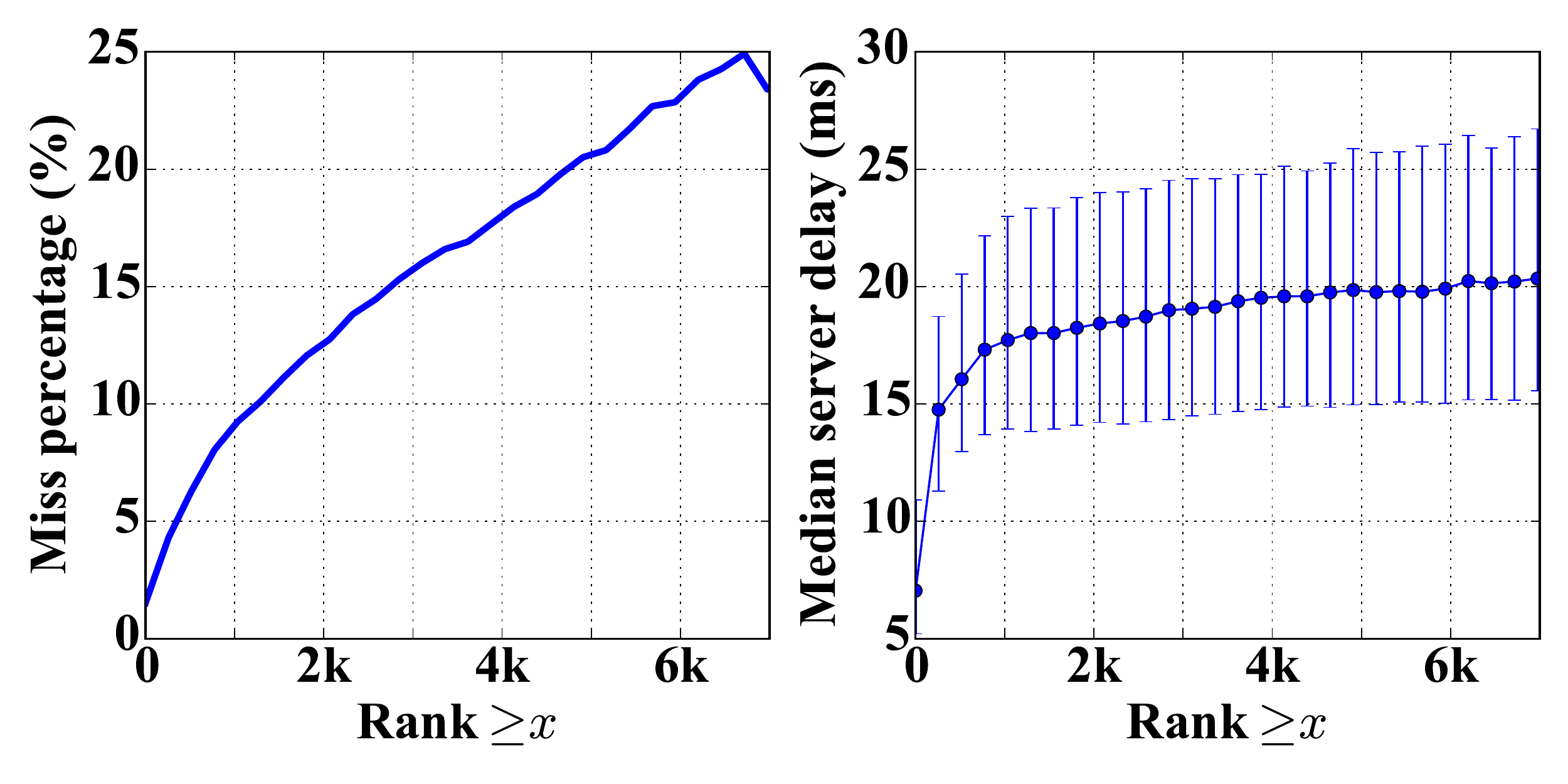}
\caption{Performance vs popularity: (a) miss rate vs rank, (b) CDN latency (excluding cache misses) vs rank.}
\label{f:perf_vs_rank}
\end{figure}

Instead, the \emph{unpopularity} of the content is a major cause of the persistent server-side problems. For less popular videos, the chunks often need to come from disk or, worse yet, the backend server. Figure~\ref{f:perf_vs_rank}(a) shows the cache miss percentage versus video rank (most popular video is ranked first) using data from one day. The cache miss ratio drastically increases for unpopular videos. Even on a cache hit, unpopular videos experience higher server delay, as shown in Figure~\ref{f:perf_vs_rank}(b).  The figure shows mean server latency after removing cache misses (i.e., no backend communication). The unpopular content generally experiences a higher latency due to higher read (seek) latency from disk. 

\noindent\textbf{Take-aways.}
First, the persistence of cache misses could be addressed by pre-fetching the subsequent chunks of a video session after the first miss. Pre-fetching of subsequent chunks would particularly help the unpopular videos since backend latency makes up a significant part of their overall latency and could be avoided; thus, to help unpopular videos the CDN server could cache the first few chunks of all videos to reduce startup delay. 

Second, when an object cannot be served from local cache, the request will be sent to the backend server. For a popular object many near-simultaneous requests may overwhelm the backend service; thus, the ATS retry timer is used to reduce the load on the backend servers, the timer introduces too much delay for cases where the content is available on local disk. Since the timer affects $35\%$ of chunks, we recommend decreasing the timer for disk accesses. 

\vspace{0.1in}\noindent\textbf{3. Load vs. performance due to cache-focused client mapping.}
We have observed that more heavily loaded servers offer \emph{lower} overall CDN latency. This result was initially surprising since we expected busier servers to have worse performance, however, this load-performance paradox is explainable by the \emph{cache-focused mapping} CDN feature: As a result of cache-based assignment of clients to CDN servers, servers with less popular content (higher rank videos) have more chunks with either higher read latency (due to the ATS retry-timer) as the content is not fresh in memory, or worse yet, need to be requested from backend due to cache-misses. 

While unpopular content leads to lower performance, because of lower demand it also produces fewer requests, hence servers that serve less popular content seem to have worse performance at a lower load.

\noindent\textbf{Take-away.}
To achieve better utilization of servers and balancing the load, in addition to cache-focused routing, popular content can be explicitly partitioned/distributed among servers. For example, given that the top 10\% of videos make up 66\% of requests, distributing only the top 10\% of popular videos across servers can balance the load.
  
\subsection{Network performance problems}
\label{s:net}
Network problems can manifest in form of increased packet losses, reordering, high latency, high variation in latency, and low throughput. Each can be persistent (e.g., far away clients from a server have persistent high latency) or transient (e.g., spike in latency caused by congestion). In this section, we characterize these problems.

Distinguishing between a transient and a persistent problem matters because a good ABR may \emph{adapt} to temporary problems (e.g., lowering bitrate), but it cannot avoid bad quality caused by persistent problems (e.g., when a peering point is heavily congested, even the lowest bitrate may see re-buffering). Instead, persistent problems require \emph{corrective actions} taken by the video provider (e.g., placement of new PoPs closer to client) or ISPs (e.g., additional peering).

We characterize the impact of loss and latency on QoE next. To characterize long-term problems, we aggregate sessions into /24 IP prefixes since most allocated blocks and BGP prefixes are /24 prefixes~\cite{Poese:2011:IGD:1971162.1971171,Freedman:2005:GLI:1251086.1251099}.

\begin{figure}[t!]
\centering
  \includegraphics[width=0.8\linewidth]{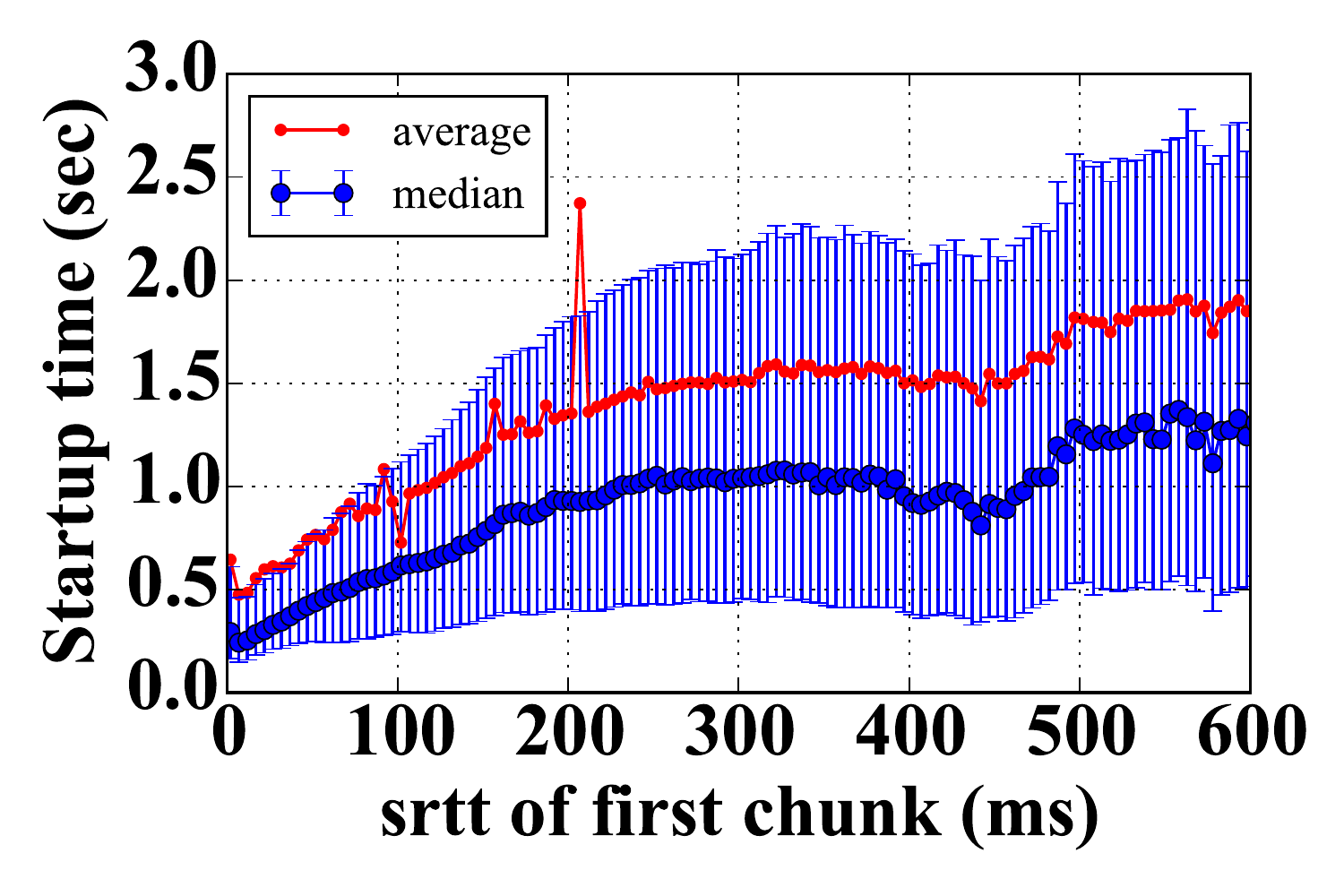}
  \caption{Average and median startup delay vs. network latency, error bars show the interquartile range (IQR). }
  \label{f:join_vs_srtt}
\end{figure}

\begin{figure}[t!]
\centering
  \includegraphics[width=0.8\linewidth]{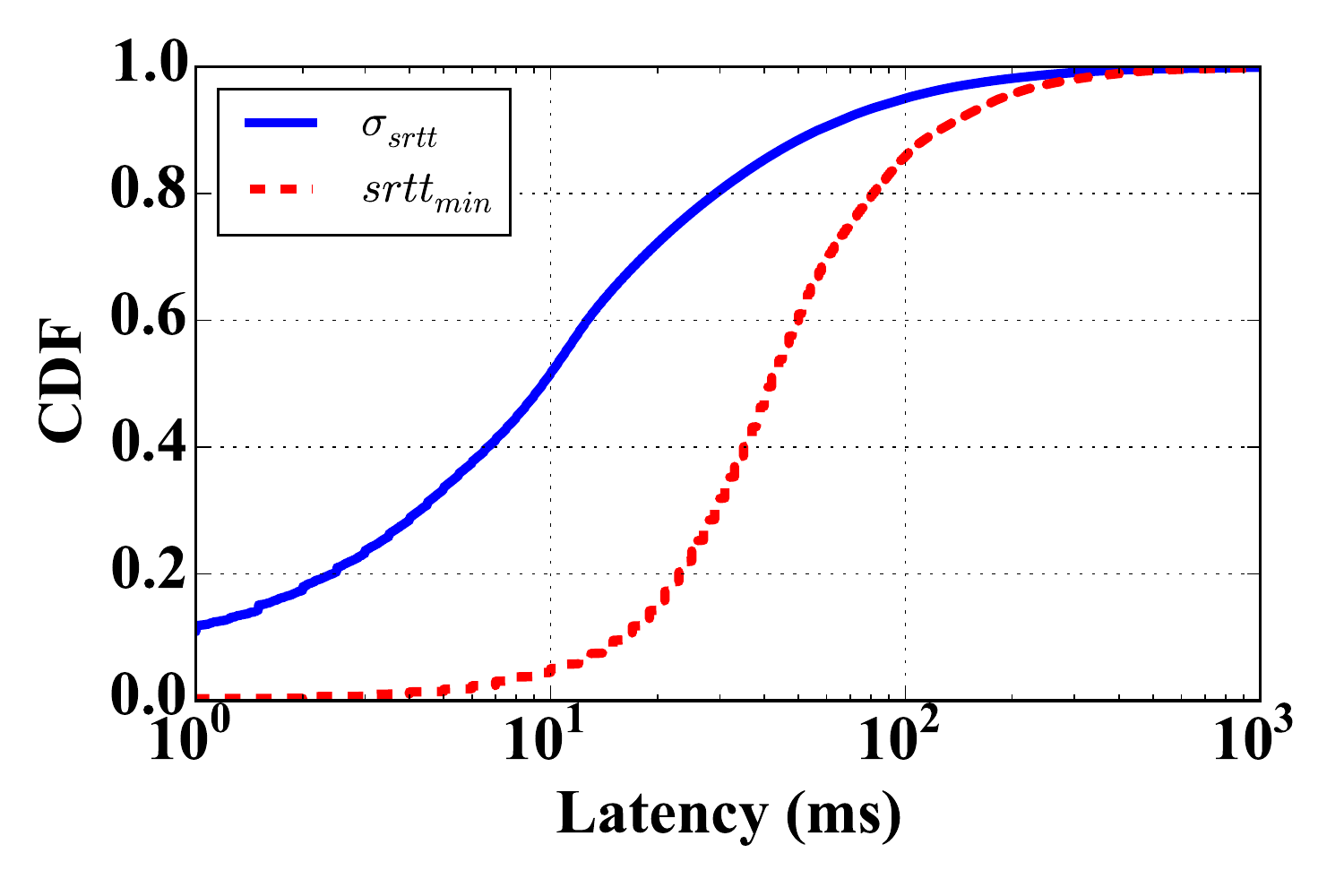}
  \caption{CDF of baseline ($srtt_{min}$) and variation in latency ($\sigma_{srtt}$) among sessions.}
  \label{f:cdf_latency_session}
\end{figure}

Figure~\ref{f:join_vs_srtt} shows the effect of network latency during the first chunk on video QoE, specifically, startup delay, across sessions. High latency in a session could be caused by a persistently high baseline (i.e., high $srtt_{min}$)\footnote{Note that TCP's estimate of RTT, SRTT, is an EWMA average; hence $srtt_{min}$ is higher than the minimum RTT seen by TCP; the bias of this estimator, however, is not expected to be significant for our study since it is averaged.}, or fluctuations in latency as a result of transient problems (i.e., high variation, $\sigma_{srtt}$). Figure~\ref{f:cdf_latency_session} depicts the CDF of both of these metrics across sessions. We see that both of these problems exist among sessions; we characterize each of these next. 

\vspace{0.1in}\noindent\textbf{1. Persistent high latency caused by distance or enterprise path problems.} 
From Figure~\ref{f:cdf_latency_session} we can see some sessions have a high minimum RTT. To analyze the minimum latency, it is important to note that the SRTT samples are taken after 500ms from the beginning of the chunk's transmission; hence, if a chunk has self-loading~\cite{Jain:2002:EAB:633025.633054}, the SRTT sample may reflect the additional queuing delay, not just the baseline latency. To filter out chunks whose SRTT may have grown while downloading, we use an estimate of the initial network round-trip time ($rtt_0$) per-chunk.
Equation~\ref{e:rtt0} shows that $D_{FB}-(D_{CDN}+D_{BE})$ can be used as an upper-bound estimation of $rtt_0$. We take the minimum of SRTT and $rtt_0$ per-chunk as the baseline sample. Next, to find the minimum RTT to a session or prefix, we take the minimum among all these per-chunk baseline samples in the session or prefix.

To discover the underlying cause of persistently high latency, we aggregate sessions into /24 client prefixes. The aggregation overcomes client last-mile problems, which may increase the latency for one session, but are not persistent problems. A prefix has more RTT samples than a session; hence, congestion is less likely to inflate all samples.

We focus our analysis on prefixes in $90^{th}$ tail of latency, where $srtt_{min} > 100ms$; which is a high latency for cable/broadband connections (note that our node footprint is largely within North America). To ensure that a temporary congestion or routing change has not affected samples of a prefix, and to understand the lasting problems in poor prefixes, we repeat this analysis for every day in our dataset and calculated the recurrence frequency, $\frac{\textit{\#days prefix in tail }}{\# days}$. We take the top 10\% of prefixes with highest re-occurrence frequency as prefixes with a persistent latency problem. This set includes $57k$ prefixes.

\begin{figure}[t!]
\centering
  \includegraphics[height=0.45\linewidth]{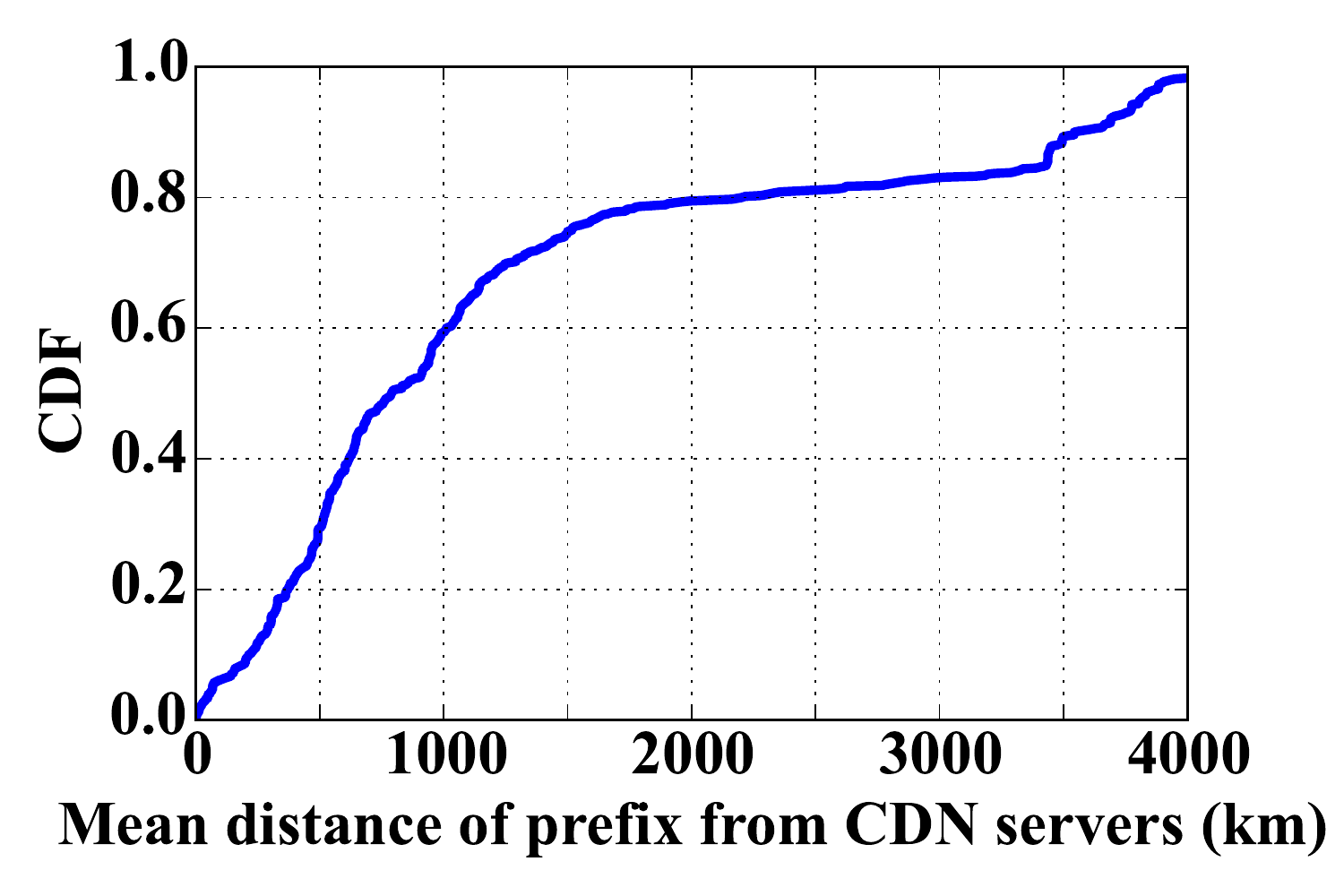}
  \caption{Mean distance (km) of US prefixes in the tail latency from CDN servers.}
  \label{f:tail_rtt_distance}
\end{figure}

While over $93\%$ of clients are located in the US, from these $57k$ prefixes 75\% are located outside the US and are spread across 96 different countries. These non-US clients are often limited by geographical distance and propagation delay.
However, among the 25\% of prefixes located in the US, \emph{the majority are close to CDN nodes.}
Since IP geolocation packages have been shown to unfairly favor a few countries, in particular the US with 45\% of entries~\cite{Poese:2011:IGD:1971162.1971171}, we focus our geo-specific analysis to US clients. Figure~\ref{f:tail_rtt_distance} shows the relationship between the $srtt_{min}$ and geographical distance of these prefixes in the US. If a prefix is spread over several cities, we use the average of their distances to the CDN server.
Among high-latency prefixes inside the US within a 4km distance, only about 10\% are served by residential ISPs while the remaining 90\% of prefixes originate from corporations and private enterprises. 

\noindent\textbf{Take-away:}
Finding clients who suffer from persistent high latency due to geographical distance helps video content providers in better placement of new CDN servers and traffic engineering. 
It is equally important to recognize close-by clients suffering from bad latency to (1) refrain from over provisioning more servers in those areas and wasting resources, and, (2) identify the IP prefixes with known persistent problems and adjust the streaming algorithm accordingly, for example, to start the streaming with a more conservative initial bitrate.

\vspace{0.1in}\noindent\textbf{2. Residential networks have lower latency variation than enterprises.}
To measure RTT variation, we calculate the coefficient of variation (CV) of SRTT in each session, which is defined as the standard deviation over the mean of SRTT. Sessions with low variability have $CV<1$ and sessions with high SRTT variability have $CV>1$. For each ISPs and organization, we measure the ratio of sessions with $CV>1$ to all sessions. We limit the result to ISPs/organizations that have least 50 video streaming sessions to provide enough evidence of persistence. Table~\ref{t:cvrtt} shows the top ISPs/organizations with highest ratio. Enterprises networks make up most of the top of this list. To compare this with residential ISPs, we analyzed 5 major residential ISPs and found that about 1\% of sessions have $CV>1$. 
Similarly, we speculate enterprise path issues cause this problem but we do not have in-network measurements to further diagnose this problem.

\begin{figure}[t!]
\centering
  \includegraphics[width=0.8\linewidth]{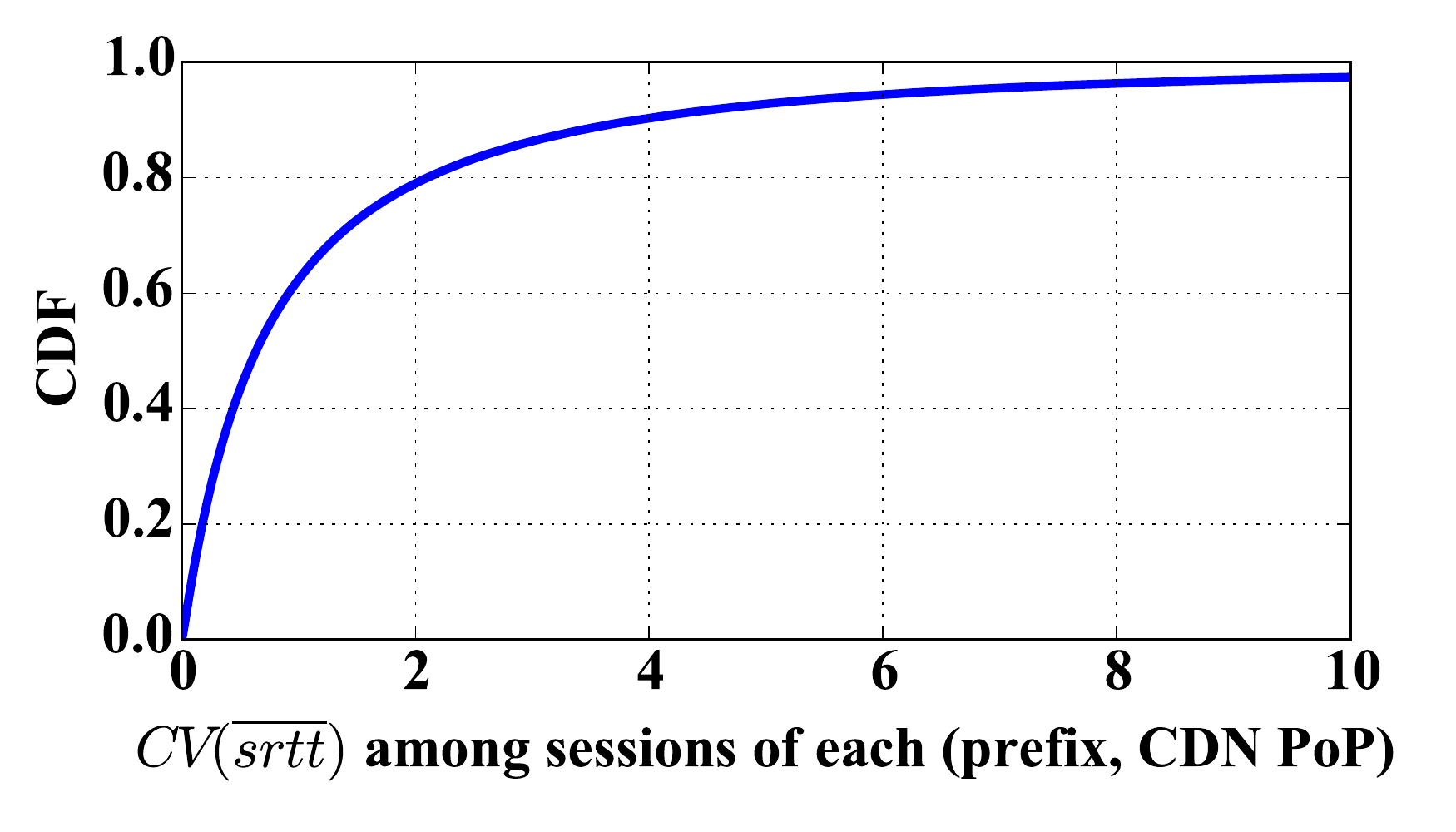}
  \caption{CDF of path latency fluctuations: CV of latency per path, a path is defined by a (prefix, PoP) pair.}
  \label{f:cv_rtt_prefix}
\end{figure}

In addition to per-session fluctuations in latency, we have characterized the fluctuations of latency in prefixes as shown in Figure~\ref{f:cv_rtt_prefix}. For this analysis, we used the average $srtt$ of each session as the sample latency. To find the coefficient of variance among all (source, destination) paths, sessions are grouped based on their prefix and CDN PoP. We can see that 40\% of (prefix, PoP) pairs belong to paths with high latency variation. ($CV > 1$)

\noindent\textbf{Take-away:} 
Recognizing which clients are more likely to suffer from latency fluctuations is valuable for content providers because it helps them make informed decisions according to clients' needs. In particular, the player bitrate adaptation and CDN traffic engineering algorithms can use this information to make better decisions resulting in a robust streaming quality despite the high fluctuations in latency. For example, the player can make more conservative bitrate choices, lower the inter-chunk wait time (i.e., request chunks sooner than usual), and increase the buffer size to deal with fluctuations.

\begin{table}\centering
\scriptsize
\ra{1.3}
\begin{tabularx}{\linewidth}{lrrr}
\textbf{isp/organization}&\textbf{\#sessions with $CV>1$}&\textbf{\#all sessions}&\textbf{Percentage} \\ \toprule
Enterprise\#1& 30 & 69 &	43.4\%\\
\hline
Enterprise\#2&	4,836 	& 11,731 & 41.2\%\\
\hline
Enterprise\#3&	1,634 & 	4,084 & 	40.0\%\\
\hline
Enterprise\#4&	83	& 208	 & 39.9\%\\
\hline
Enterprise\#5 &	81 & 203	& 39.9\%\\
\hline
\bottomrule
\end{tabularx}
\caption{ISP/Organizations with highest percentage of sessions with $CV(SRTT)>1$.}
\label{t:cvrtt}
\end{table}

\begin{figure}[t!]
\centering
\subfigure[CDF of session length with and without loss]{ \includegraphics[width=0.8\linewidth]{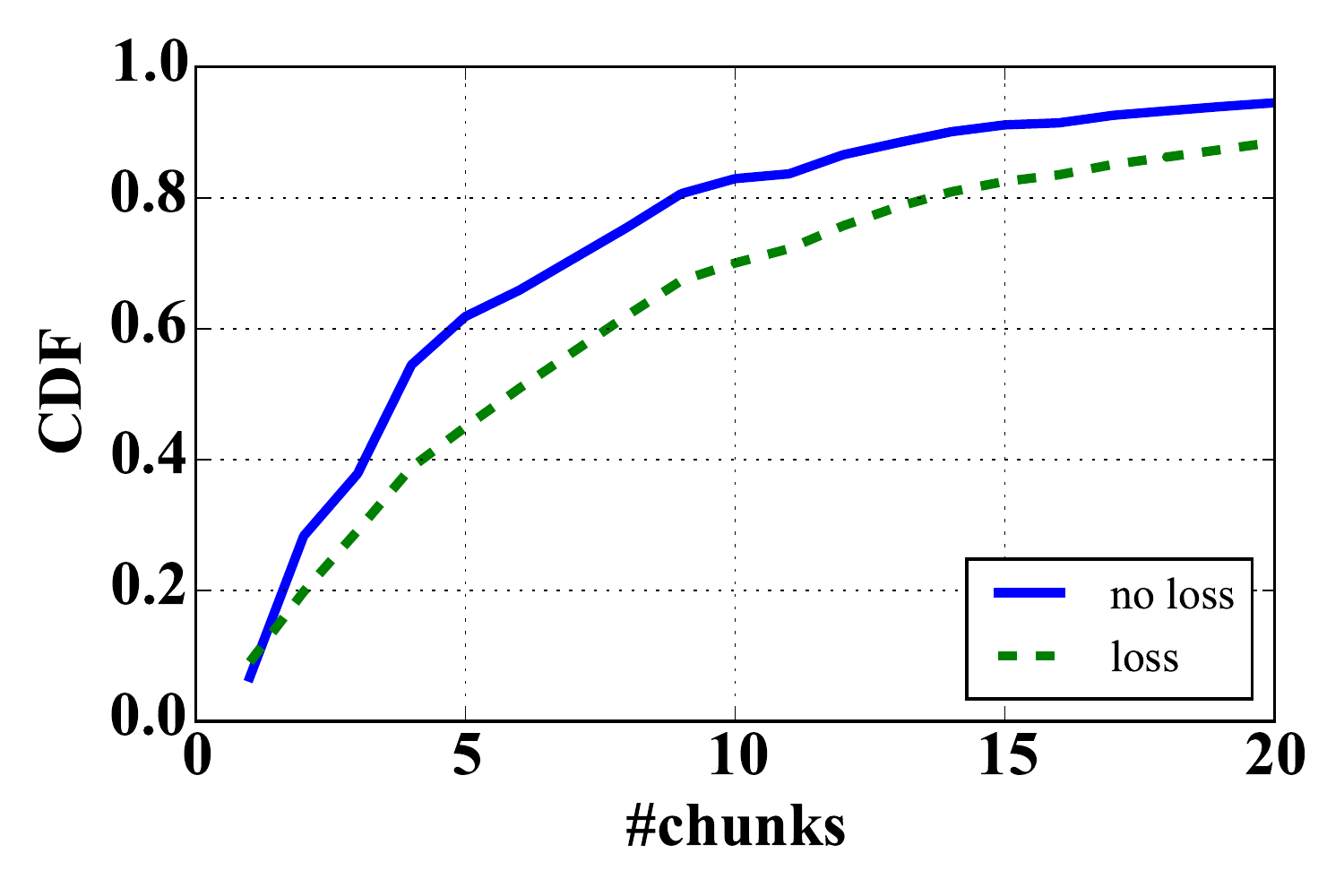}}

\subfigure[CDF of Average bitrate with and without loss]{ \includegraphics[width=0.8\linewidth]{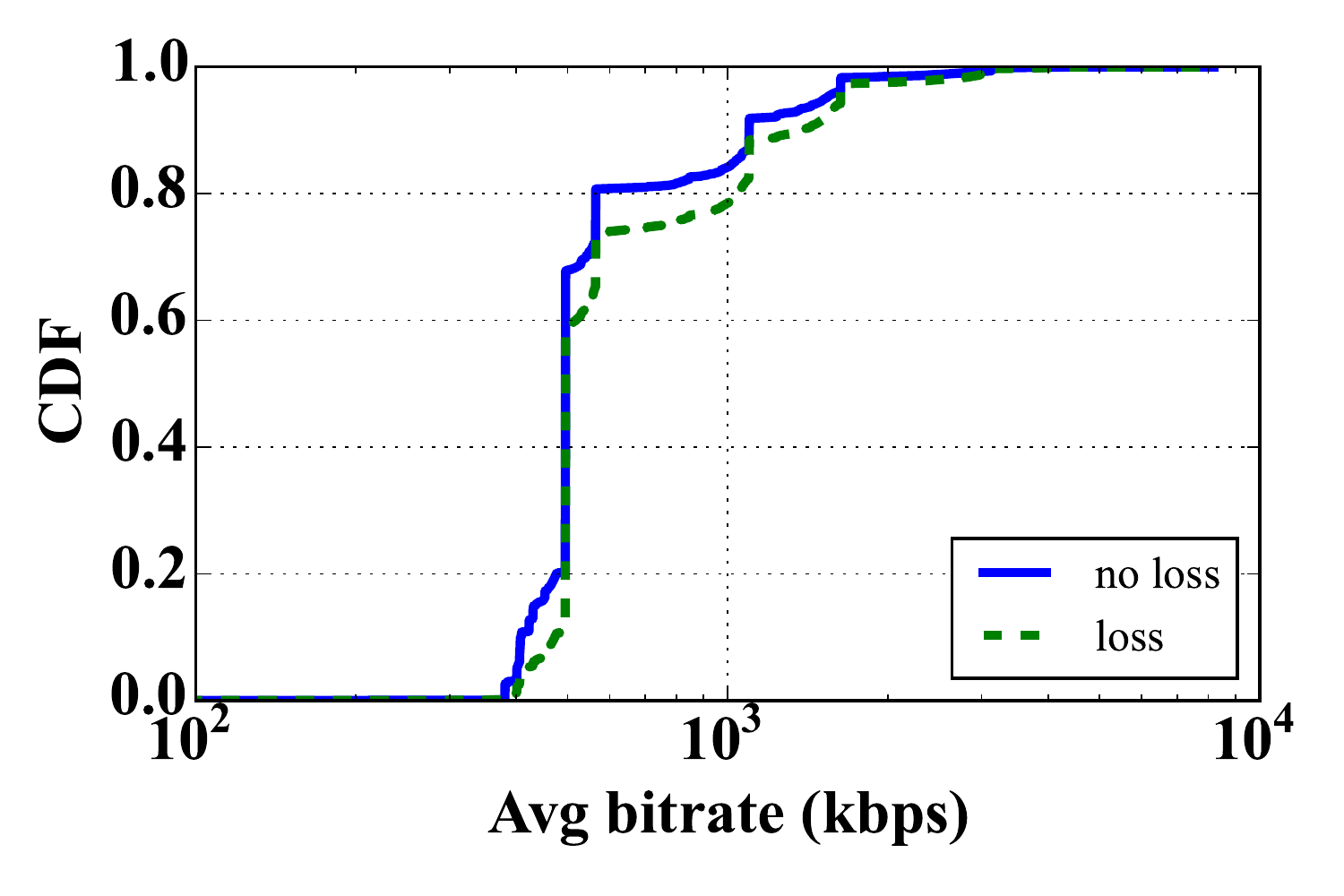}}

\subfigure[CCDF (1-CDF) of Re-buffering rate with and without loss]{ \includegraphics[width=0.8\linewidth]{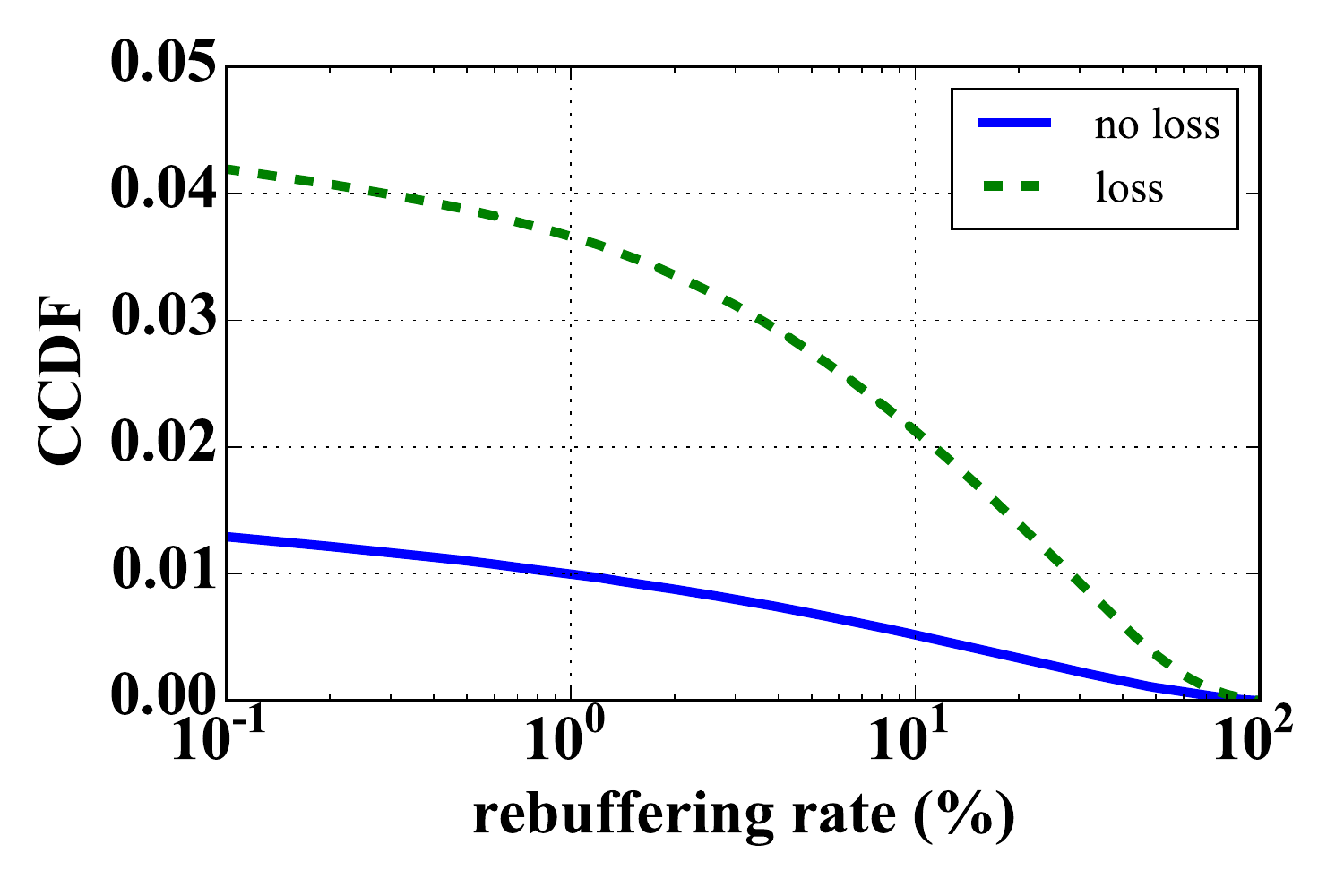}}
\caption{Differences in session length, quality, and re-buffering with and without loss.}
\label{f:namework}
\end{figure}

\vspace{0.1in}\noindent\textbf{3. Earlier packet losses have higher impact on QoE.}
We use the retransmission counter as an indirect way to study the effect of packet losses. The majority of the sessions ($>90\%$) have 
a retransmission rate of less than 10\%, with 40\% of sessions experiencing no loss. While 10\% is a huge loss rate in TCP, not every retransmission is caused by an actual loss (e.g., early retransmit optimizations, underestimating RTO, etc.).
Figure~\ref{f:namework} shows the differences between sessions with and without loss in three aspects: (a) number of chunks (are these sessions shorter?), (b) bitrate (similar quality?), and (c) re-buffering. Based on this figure, we can conclude that the session length and bitrate distributions are almost similar between the two groups, however, re-buffering difference is significant and sessions without loss have better QoE. 

\begin{figure}[t!]
\centering
  \includegraphics[width=0.8\linewidth]{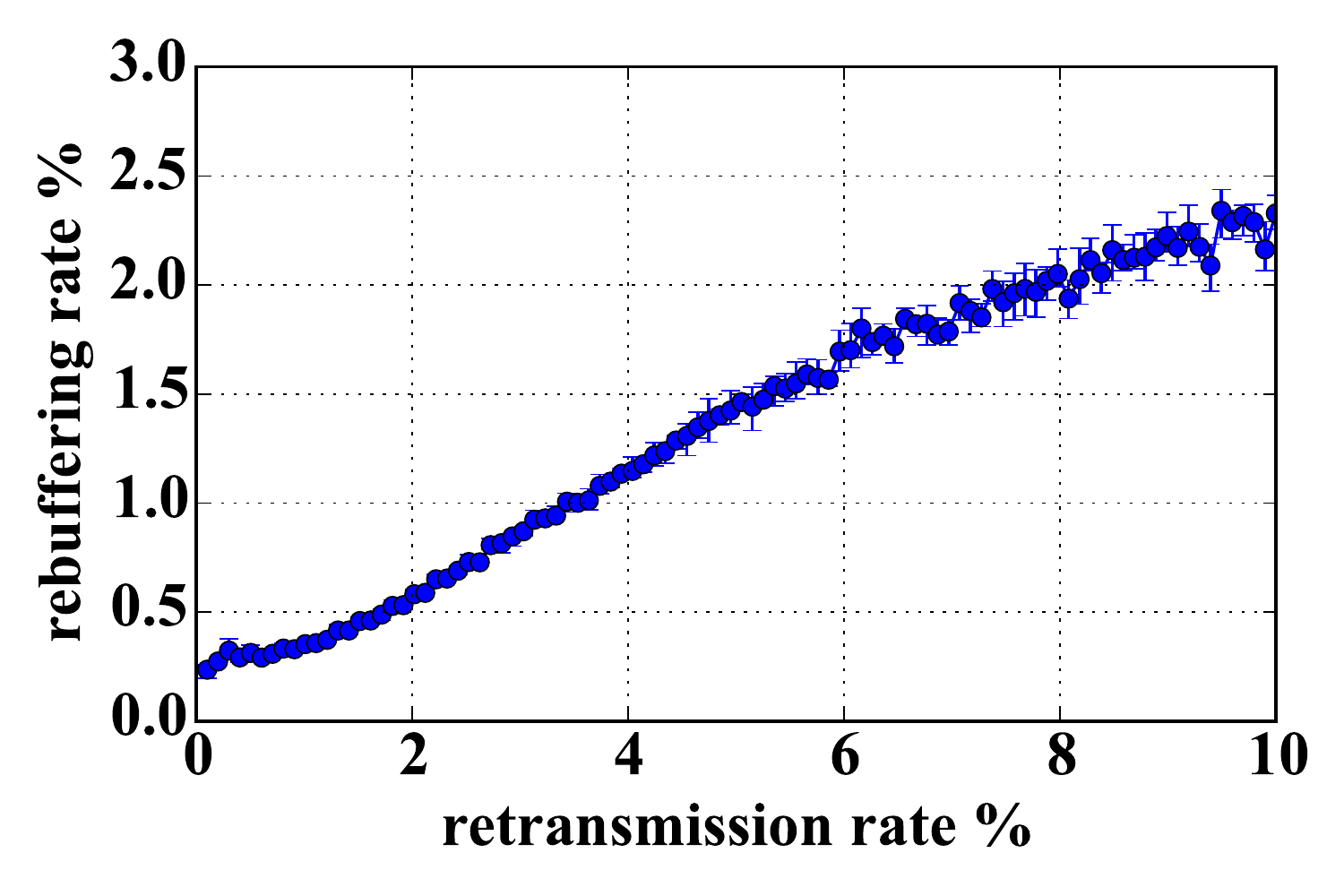}
  \caption{Rebuffering vs retransmission rate in sessions.}
  \label{f:session_loss}
\end{figure}

\begin{figure}
\centering
\includegraphics[width=0.8\linewidth]{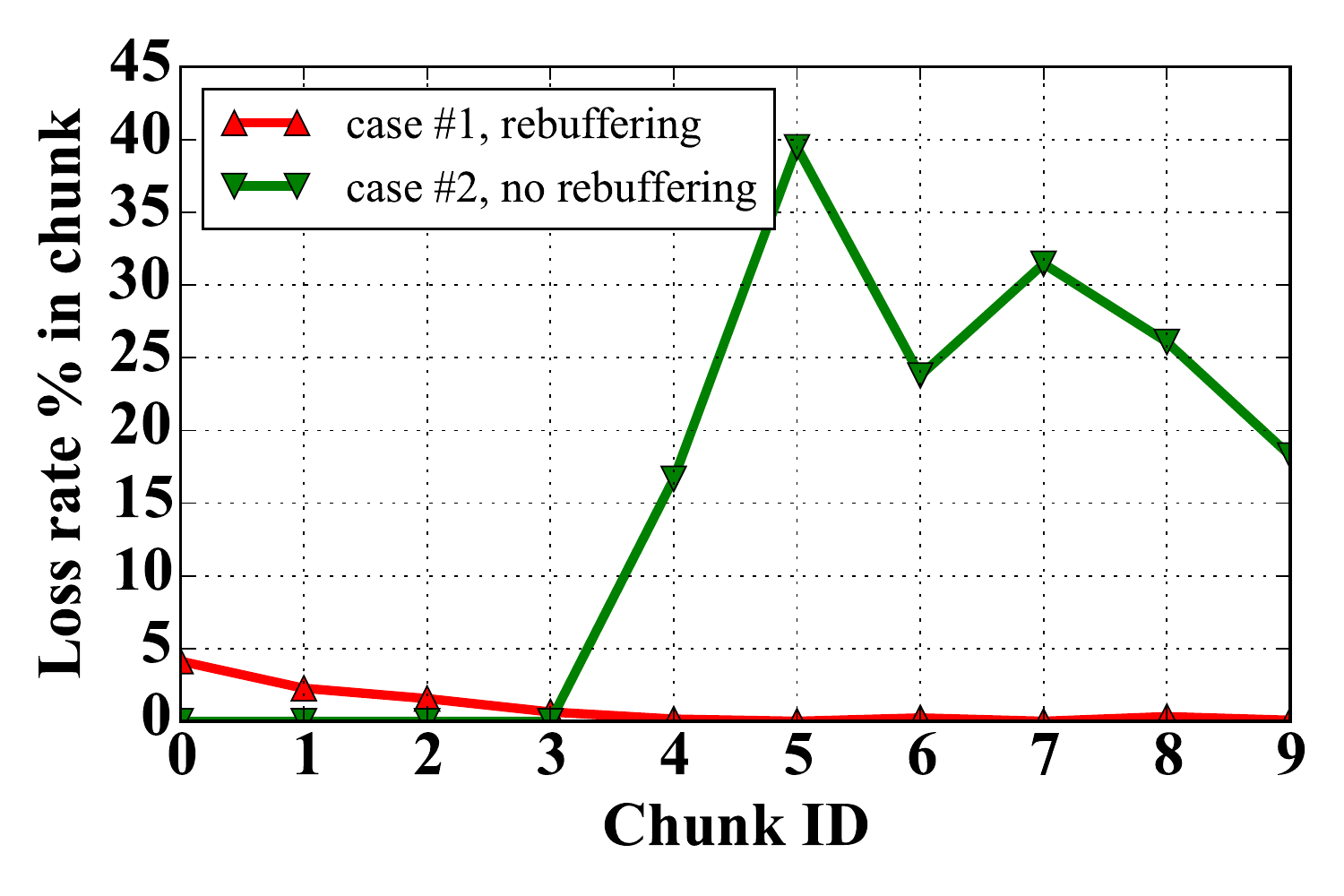}
\caption{Example case for loss vs QoE.}
\label{f:loss:a}
\end{figure}

While higher loss rates generally indicate higher re-buffering (Figure~\ref{f:session_loss}), the loss rate of a TCP connection does not necessarily correlate with its QoE, rather, time of loss matters too. Figure~\ref{f:loss:a} shows two example cases, both sessions have 10 chunks with similar bitrates, cache statuses, and SRTT samples. Case \#1 has a retransmission rate of is 0.75\%, compared to 22\% in case \#2; but it experiences dropped frames and re-buffering despite the lower loss rate. As Figure ~\ref{f:loss:a} shows, the majority of losses in case \#1 happen in the first chunk whereas case \#2 has no loss during the first four chunks, building up its buffer to 29.8 seconds before a loss happens and successfully avoids re-buffering.

\begin{figure}[t!]
\centering
  \includegraphics[width=0.8\linewidth]{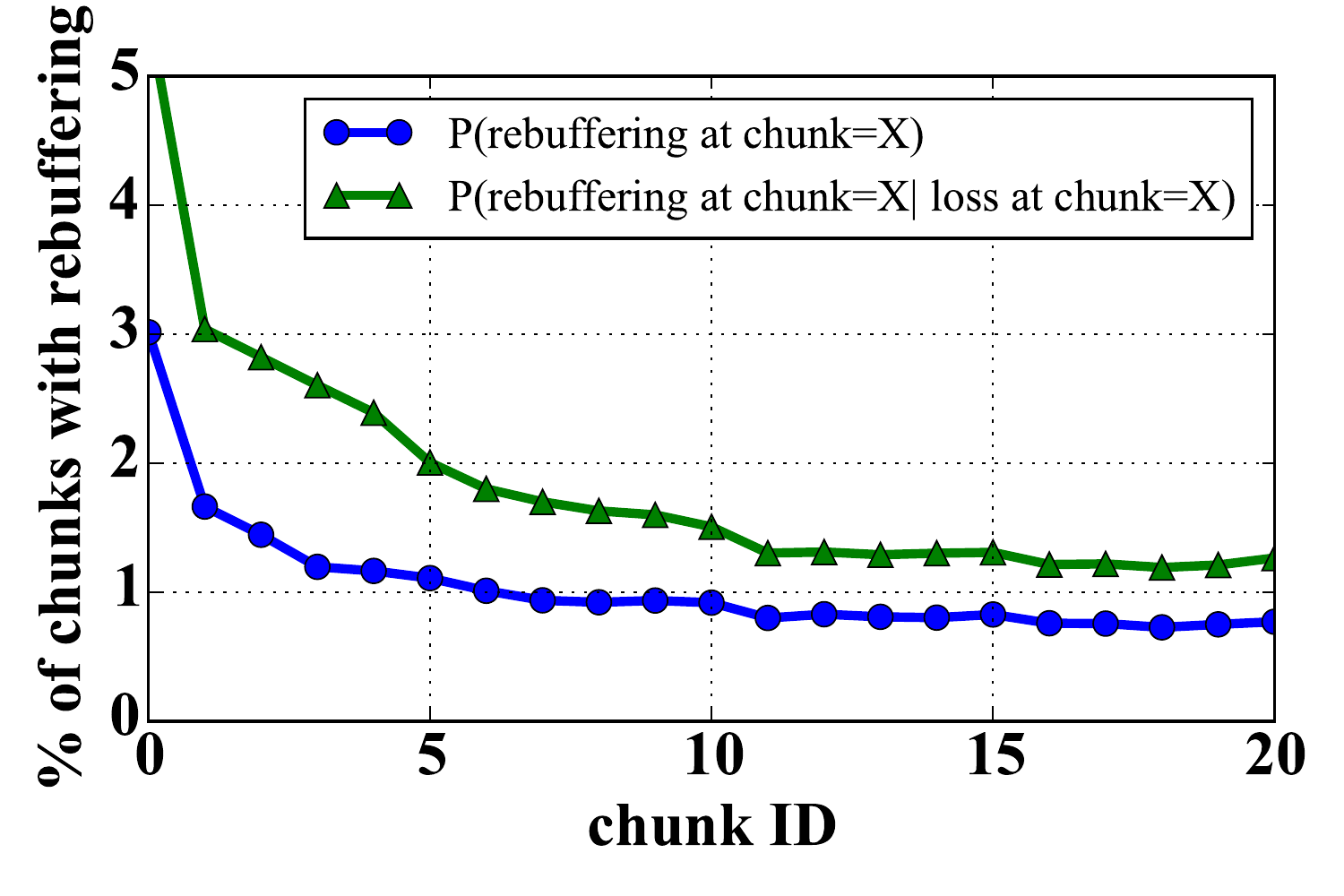}
  \caption{Re-buffering frequency per chunkID, Re-bufffering frequency given loss per chunkID.}
  \label{f:rebuf_loss}
\end{figure}

Because the buffer can hide the effect of subsequent loss, we believe it is important to not just measure loss rate in video sessions, but also \emph{the chunkID that experiences loss}. Loss during earlier chunks has more impact on QoE because the playback buffer holds less data. We expect losses during the first chunk to have the worst effect on re-buffering. Figure~\ref{f:rebuf_loss} shows two studies; (1) P($rebuf at chunk$=X), which is the percentage of chunks with that chunk ID that had a re-buffering event; and (2) P(re-buffering at chunk=X$|$loss at chunk$=X)$, which is the same percentage conditioned on occurrence of a loss during the chunk. While occurrence of a loss in any chunk increases the likelihood of a re-buffering event, this increase is more significant for the first chunk. 

\begin{figure}[t!]
\centering
  \includegraphics[width=0.8\linewidth]{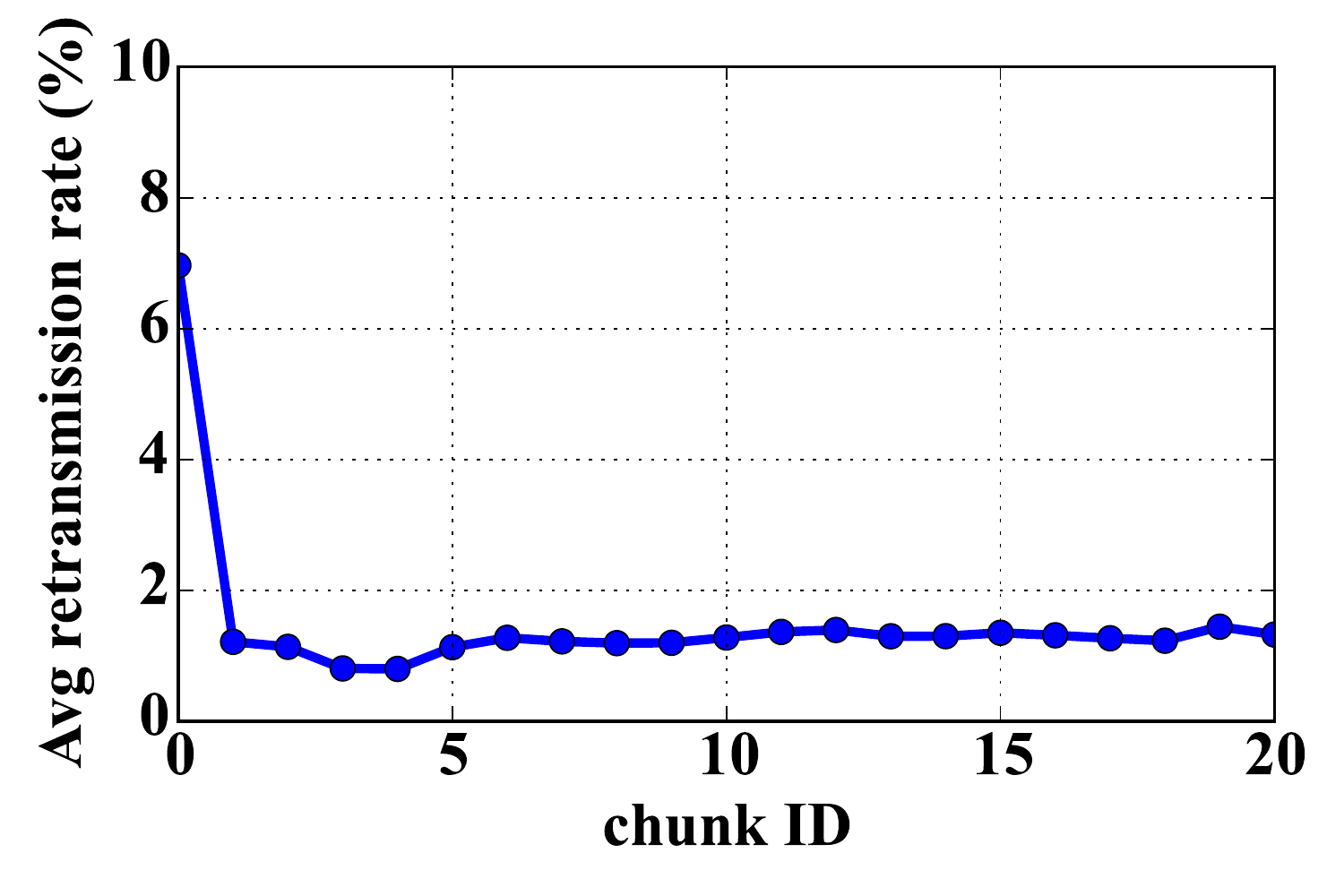}
  \caption{Average per-chunk retransmission rate.}
  \label{f:avg_loss_id}
\end{figure}

To make matters worse, we observed that losses are more likely to happen on the first chunk: Figure~\ref{f:avg_loss_id}  shows the average per-chunk retransmission rate is. The bursty nature of TCP losses towards the end of slow start~\cite{832483} could be the cause of higher loss rates during the first chunk, which is avoided in subsequent chunks after transitioning into congestion avoidance state.

\noindent\textbf{Take-aways:}
While measuring a TCP connection's loss rate is a widely used approach in estimating application's perceived performance, due to the existence of buffer in video streaming applications, the session-wide loss rate does not necessarily correlate with QoE; rather, the relative location of loss in session matters too. Earlier losses impact QoE more, with the first chunk having the biggest impact.

Due to the bursty nature of packet losses in TCP slow start caused by the exponential growth, the first chunk has the highest per-chunk retransmission rate. We suggest server-side pacing solutions~\cite{Ghobadi:2012:TRL:2342821.2342838} to work around this issue.

\vspace{0.1in}\noindent\textbf{4. Throughput is a bigger problem than latency.}
To separate chunks based on overall performance, we use the following intuition:  
\emph{The playback buffer decreases when it takes longer to download a chunk than there are seconds of video in the chunk.} This insight can be summarized as the following formula to detect chunks with bad performance when score is less than 1 ($\tau$ is the chunk duration):
\begin{equation}
perf_{score} = \frac{\tau}{D_{FB} + D_{LB}}
\end{equation}

We use $D_{LB}$ as a ``measure'' of throughput. Both latency ($D_{FB}$) and throughout ($D_{LB}$) play a role in this score. We observed that while the chunks with bad performance have generally higher latency and lower throughput than chunks with good performance, throughput is a more defining metric in overall performance of the chunk. We define the latency share in performance by $\frac{D_{FB}}{D_{FB}+D_{LB}}$ and the throughput share by $\frac{D_{LB}}{D_{FB}+D_{LB}}$. 

\begin{figure*}
\centering
\subfigure[]{\label{f:ls:a} \includegraphics[width=0.3\linewidth]{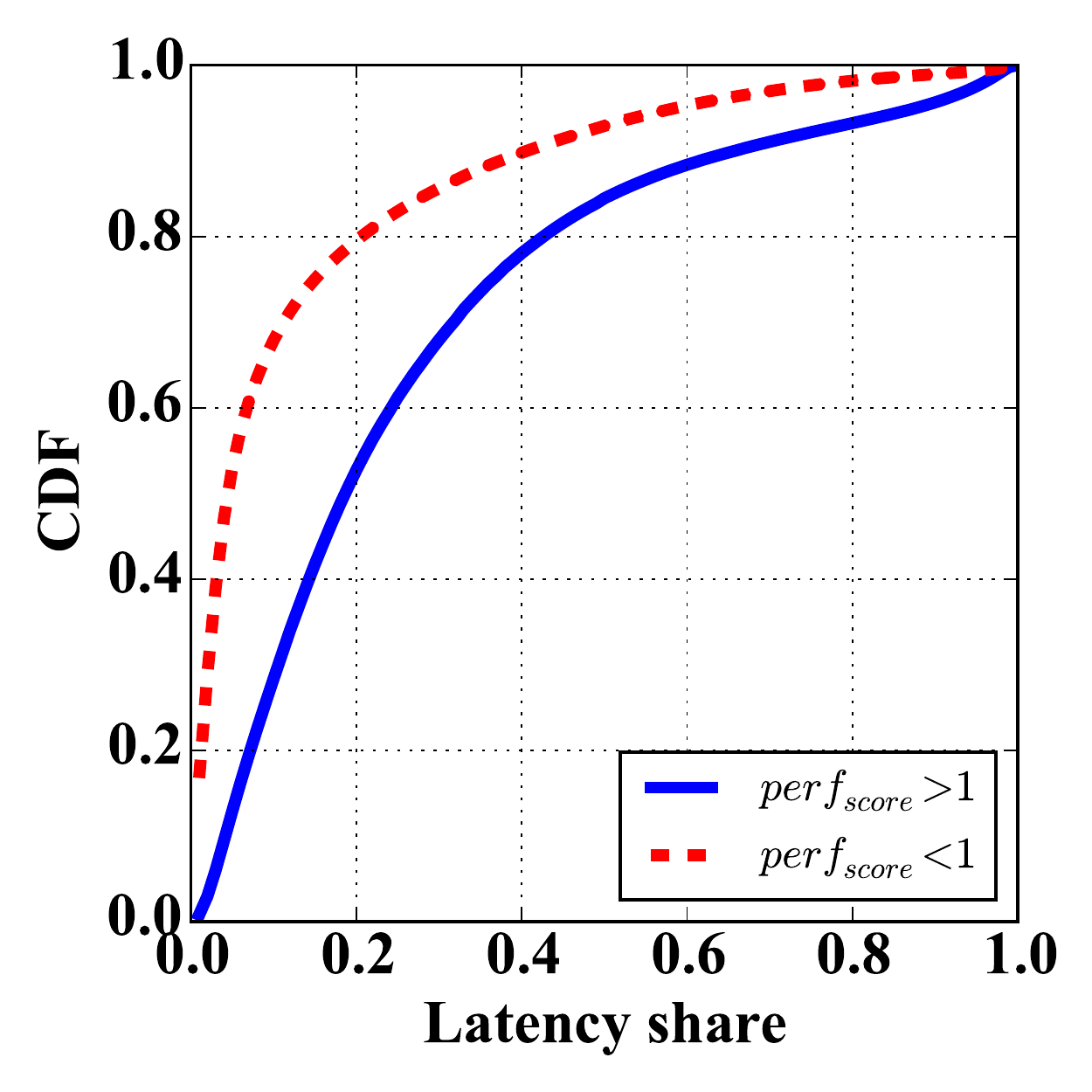}}
\subfigure[]{\label{f:ls:b} \includegraphics[width=0.3\linewidth]{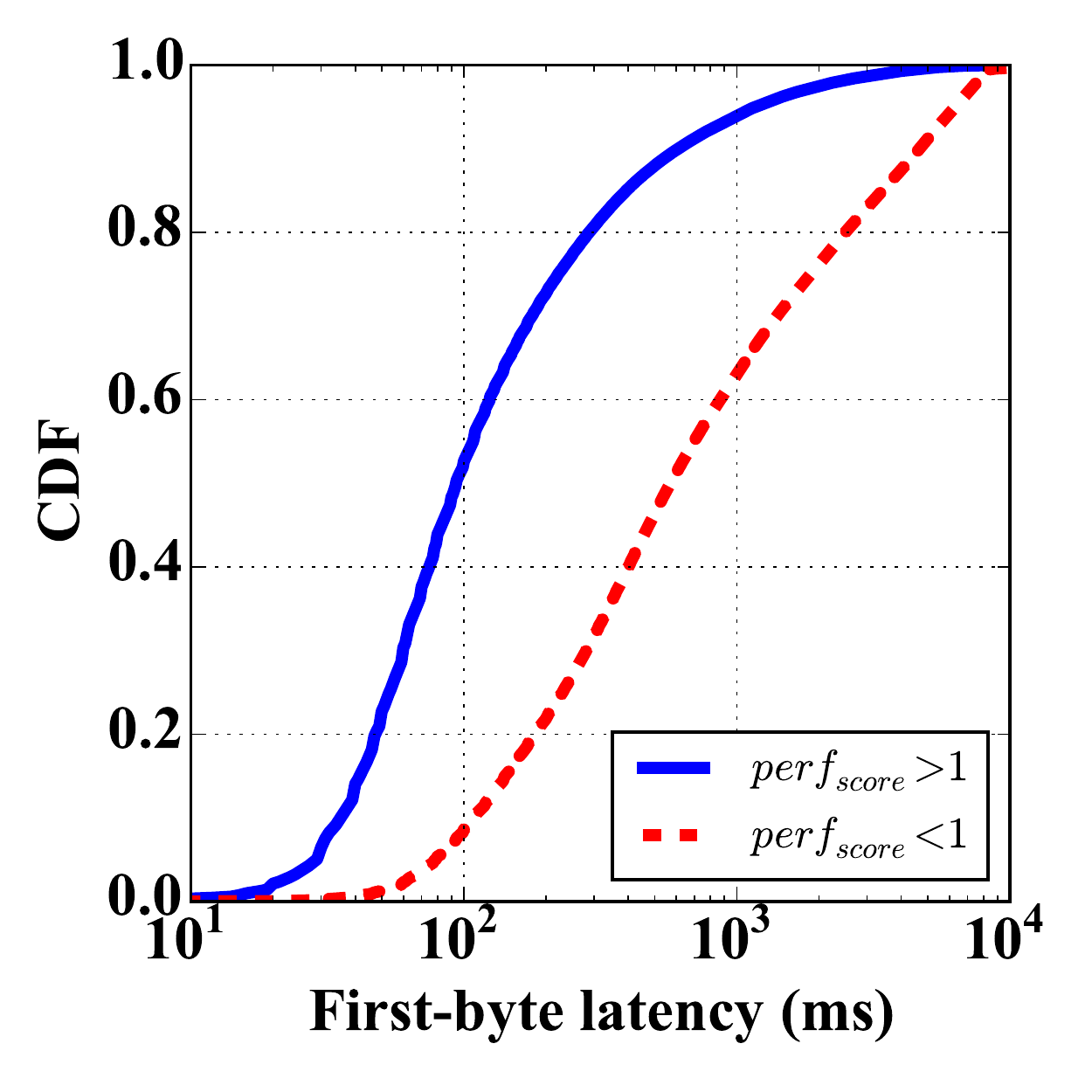}}
\subfigure[]{\label{f:ls:c} \includegraphics[width=0.3\linewidth]{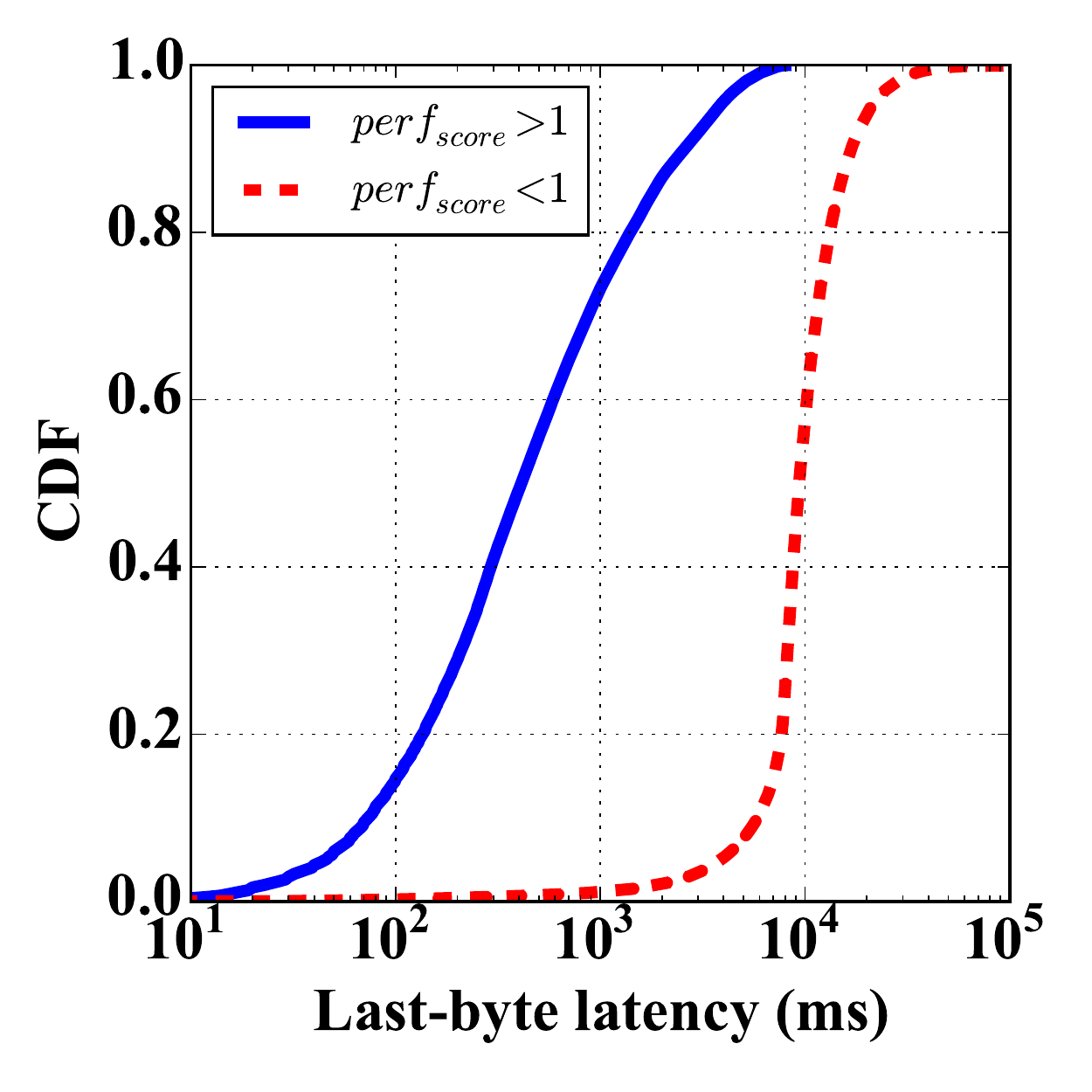}}
\caption{Latency vs throughput: (a) Latency share ($\frac{D_{FB}}{D_{FB}+D_{LB}}$), (b) $D_{FB}$, and (c) $D_{LB}$ vs performance score. }
\end{figure*}

Figure~\ref{f:ls:a} shows that that chunks with good performance generally have higher share of latency and lower share of throughput than chunks with bad performance. 
Figure~\ref{f:ls:b} shows the difference in pure values of $D_{FB}$, and Figure~\ref{f:ls:c} shows the difference in pure values of $D_{LB}$. 

While chunks with bad performance do generally have higher first and last byte delays, difference in $D_{FB}$ is negligible compared to the values of $D_{LB}$. We can see that most chunks with bad performance are limited by throughout and have a higher throughput share.

\noindent\textbf{Take-away:}
This is good news for ISPs because throughput is an easier problem to fix (e.g., establish better peering points) than latency~\cite{latency}.

\subsection{Client's Download Stack}
\noindent\textbf{1. Some chunks have a significant download stack latency.}
Video packets traversing the client's download stack (OS, browser, and the Flash plugin) may get delayed due to buffered delivery. 
In an extreme scenario, all the chunk bytes may be buffered and delivered late but all at once to the player, resulting in a huge increase in $D_{FB}$. Since the buffered data gets delivered at once or in short time windows, the \emph{instantaneous throughput} ($TP_{inst} = \frac{subchunk size}{D_{LB}}$) seems much higher at the player than the arrival rate of data from the network. We use TCP variables to estimate the download throughout per-chunk:
\begin{equation}
\textit{connection's TP } = MSS(\frac{CWND}{SRTT})
\label{e:tp}
\end{equation}

To detect chunks with this issue, we use a statistical approach to screen for outliers using standard deviation:
when a chunk is buffered in the download stack, its $D_{FB}$ is abnormally higher than
the rest (more than $2 \cdot \sigma$ greater than the mean) despite other similar latency metrics (i.e., network and server-side latency are within one $\sigma$ of the mean). 
Also, its $TP_{inst} $ is abnormally higher (more than $2 \cdot \sigma$ greater than the mean) due to the buffered data being delivered in a shorter time, while the measured connection's throughput from server (using CWND and SRTT) does not explain the increase in throughput. Equation~\ref{e:ds_std} summarizes the detection methodology:

\begin{equation}
\begin{split}
D_{FB_i} > \mu_{D_{FB}} + 2 \cdot \sigma_{D_{FB}} \\ 
TP_{{inst}_i} > \mu_{TP_{inst}} + 2 \cdot \sigma_{TP_{inst}} \\
SRTT, D_{server}, CWND < \mu + \sigma
\end{split}
\label{e:ds_std}
\end{equation}

\begin{figure}[!t]
\centering
\subfigure[First-byte delay and its constituents in the session ]{\label{f:case_ds:a}\includegraphics[width=0.8\linewidth]{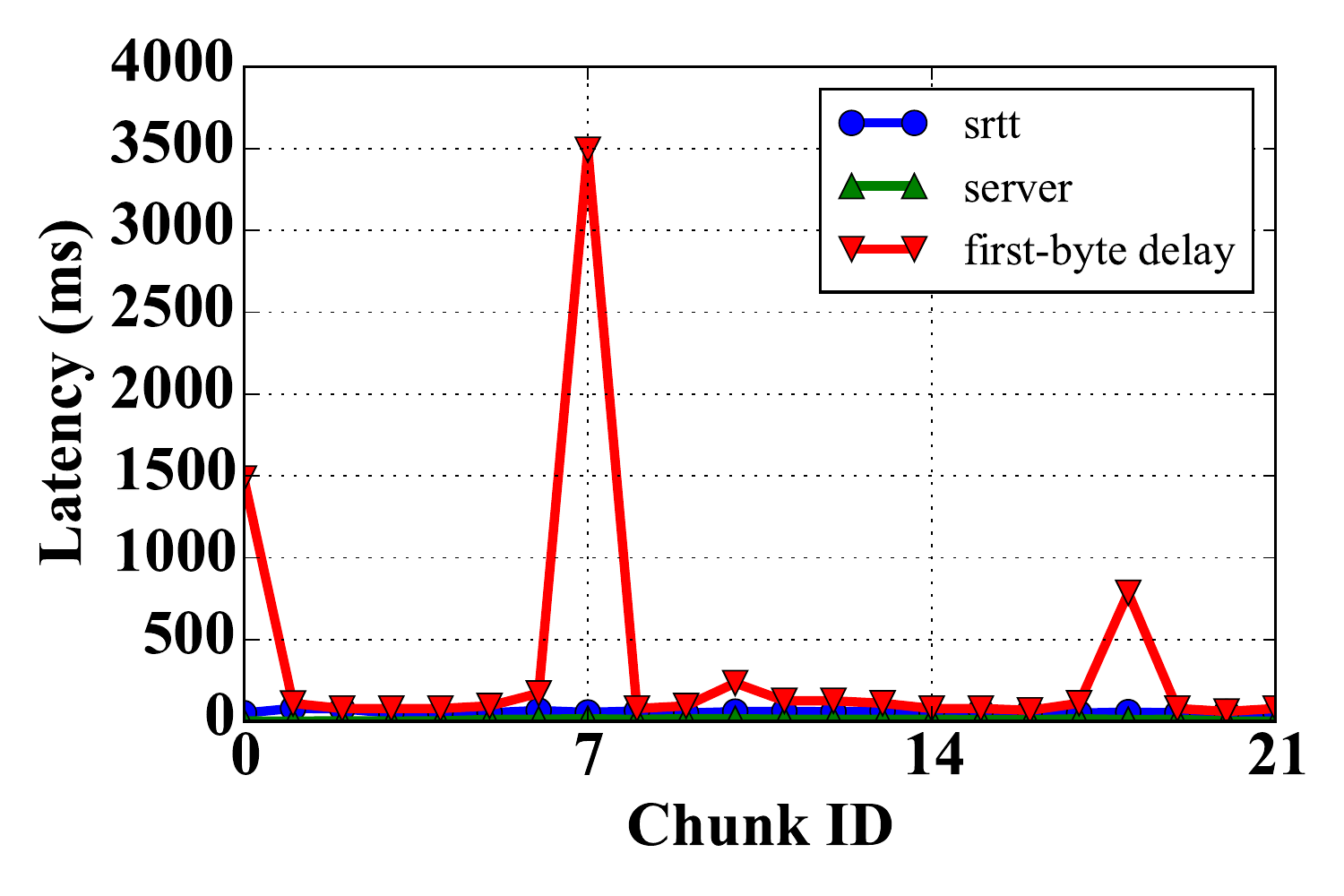}}
\subfigure[Network throughput vs instantaneous download throughput of the chunk]{\label{f:case_ds:b}\includegraphics[width=0.8\linewidth]{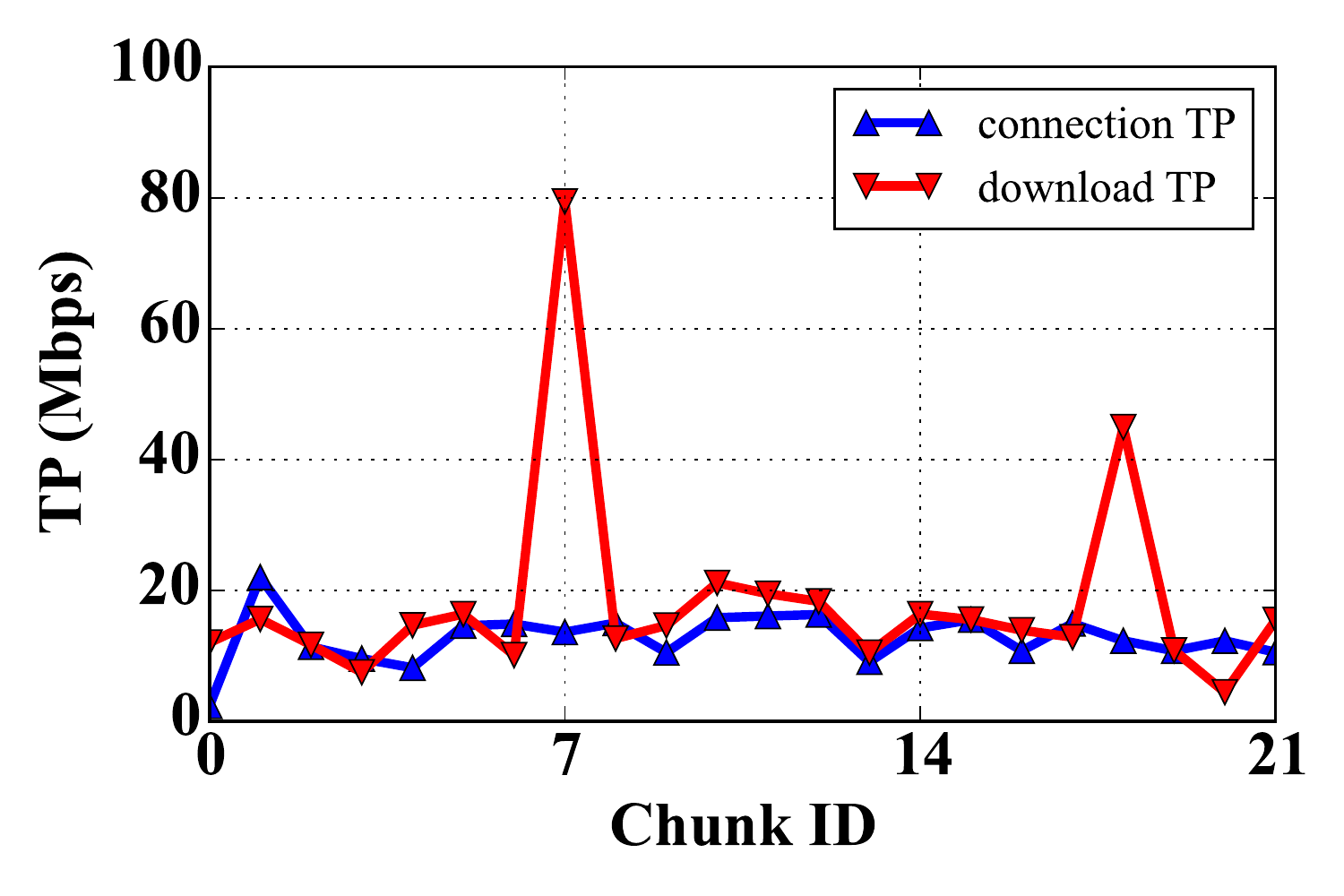}}
\caption{A case study showing the effects of client download stack (chunk\#7). }
\label{f:case_ds}
\end{figure}

Figure~\ref{f:case_ds} shows an example session that experiences the download stack problem ($DS$) taken from our dataset; our algorithm has detected chunk 7 with abnormally higher $D_{FB}$ and $TP_{inst}$ than the mean. Figure~\ref{f:case_ds:a} shows $D_{FB}$ of chunks and its constituents parts. 
We can see that the increase in chunk 7's $D_{FB}$ is not caused by a spike in backend, CDN, or network RTT. Figure~\ref{f:case_ds:b} shows that this chunk also has an abnormally high throughput that seems impossible based on the measured network throughput (Equation~\ref{e:tp}) at the server-side. These two effects combined in one chunk suggests that the chunk was delivered to the client at a normal network rate and within normal latency; but it was buffered inside the client's stack and delivered late to the player. The extra delay has helped build up more buffered data as the packets arrived from the network, leading to the highest instantaneous throughput among chunks (or lowest $D_{LB}$).

We have detected 1.7m chunks (0.32\% of all chunks) using this method that demonstrate how the client download stack can at times buffer the data and hurt performance. About 1.6m video sessions have at least one such chunk (3.1\% of sessions). 

\noindent\textbf{Take-aways:}
The download stack problem is an example where looking at one-side of measurements (CDN or client) alone would lead to wrong conclusions, where both sides may blame the network. It is only with end-to-end instrumentation that this problem can be localized.  Failing to correctly recognizing the effect of download stack on latency may lead to the following problems:

\vspace{0.1in}\noindent\emph{Over-shooting}: Some ABR algorithms use the throughput in bitrate selection process (e.g., a moving average of previous N chunks' throughput). Failing to recognize a chunk's instantaneous throughput has been affected by download stack buffering can lead to overestimation of network throughput. 

\vspace{0.1in}\noindent\emph{Under-shooting}: If the ABR algorithms are either latency-sensitive, or use the average throughput (as opposed to the instantaneous TP), the affected chunks cause underestimation in the connection's throughput. 

\vspace{0.1in}\noindent\emph{Wasting resources:} When users have seemingly low connection throughput or high latency, content providers may take a corrective decision (e.g., re-routing the client). If the download stack latency is not correctly detected, clients may be falsely re-routed (due to the seemingly high network latency from server-side), which will only waste the resources as the bottleneck is in the client's machine. 

We make two recommendations for content providers to deal with this issues: 
(1) In design of rate-based ABR algorithms that rely on throughput or latency measurements, server-side measurements (CWND, SRTT) reflect the state of the network more accurately than client-side measurements (e.g., download time) and
(2) If it is not possible to incorporate direct network measurements, the current ABR algorithms that rely on client-side measurements should exclude these outliers in their throughput/latency estimations. 
 
\vspace{0.1in}\noindent\textbf{2. Persistent download-stack problems.}
The underlying assumption in the above method is that the majority of chunks will not be buffered by download stack, hence we can \emph{detect the outlier} chunks. However, when a persistent problem in client's download stack affects all or most chunks, this method cannot detect the problem. 
If we could directly observe $rtt_0$, we would find $D_{DS}$ using Equation~\ref{e:rtt0} per-chunk. Unfortunately, the Linux kernel does not expose individual RTT samples via the \emph{tcp-info} and collecting packet traces is unfeasible in production settings.

To work with this limitation, we use a conservative estimate of the $rtt_0$ using the TCP retransmission timer RTO, as calculated by Linux kernel~\footnote{$RTO = 200ms + srtt + 4.srttvar$ according to RFC 2988~\cite{rfc2988}.}. $RTO$ is how long the sender waits for a packet's acknowledgment before deciding it is lost; hence RTO is a conservative estimate of $rtt_0$. We use RTO to estimate the lower-bound of download stack latency per-chunk:
\begin{equation}
D_{DS} \geq D_{FB} - D_{CDN} - D_{BE} - RTO
\label{e:ds_rto}
\end{equation}
Using this method, we found that in our dataset, 17.6\% of all chunks experience a 
non-zero download stack latency. In 84\% of chunks with a non-zero DS, DS share in $D_{FB}$ is higher than network and server latency, making it the major bottleneck in $D_{FB}$.
Table~\ref{t:os_br_pop} shows the top OS/browser combinations with highest persistent download stack latency. We can see that among major browsers, Safari on non-Macintosh environments has the 
highest average download stack latency. In addition, we further broke down the ``other" category and found that some less-popular browsers on Windows, in particular, Yandex and SeaMonkey, have higher download stack latencies. 
Recognizing the lasting effect of client's machine on QoE helps content providers avoid actions caused by wrong diagnosis (e.g., re-routing clients due to ~\emph{seemingly} high network latency when problem is in download stack). 

\begin{table}[!t]
\centering
\scriptsize
\rowcolors{2}{gray!25}{white}
\begin{tabularx}{\linewidth}{@{}p{1.7cm}p{1cm}p{1cm}p{1.2cm}p{1.2cm}p{1cm}@{}}\toprule
& \emph{Safari on Linux} & \emph{Safari on Windows} & \emph{Firefox on Windows} & \emph{Other on Windows} & \emph{Firefox on Mac} \\\midrule
mean DS(ms) & 1041 & 1028 & 283 & 281 & 275 \\
\hline
\end{tabularx}
\caption{OS/browser with highest $D_{DS}$.}
\label{t:os_br_pop}
\end{table}

\textbf{Take-aways and QoE impact:}
Download stack problems are worse for sessions with re-buffering: We observed that among sessions with no re-buffering, the average $D_{DS}$ is less than 100ms. In sessions with up to 10\% re-buffering, the average $D_{DS}$ grows up to 250ms, and in sessions with more than 10\% re-buffering rate, the average $D_{DS}$ is more than 500ms. Although the download stack latency is not a frequent problem, it is important to note that when it is an issue, it is often the major bottleneck in latency.

It is important to know that some machine setups (e.g., Yandex or Safari on Windows) are more likely to have persistent download stack problems. The ABR algorithm can be adjusted to use this knowledge in screening for outliers in throughput or latency estimations.

\vspace{0.1in}\noindent\textbf{3. First chunks have higher download stack latency.}
We observed that $D_{FB}$ in first chunks has a higher distribution; for example, the median $D_{FB}$ is $300ms$ higher than other chunks. 
Using packet traces and developer tools on browsers, we confirmed that this effect is not visible in OS or browser timestamps. Thus, we believe the observed difference is due to higher download stack latency of first chunk. To test our hypothesis, from all the chunks we have selected a set of \emph{performance-equivalent }chunks with the following conditions:
\noindent(1) No packet loss, (2) CWND $>$ IW (10 MSS), (3) No queuing delay, and similar SRTT (we use $60ms<SRTT<65ms$ for presentation), and (4) $D_{CDN}<5ms$, and cache-hit .

\begin{figure}[!t]
\centering
\includegraphics[width=0.8\linewidth]{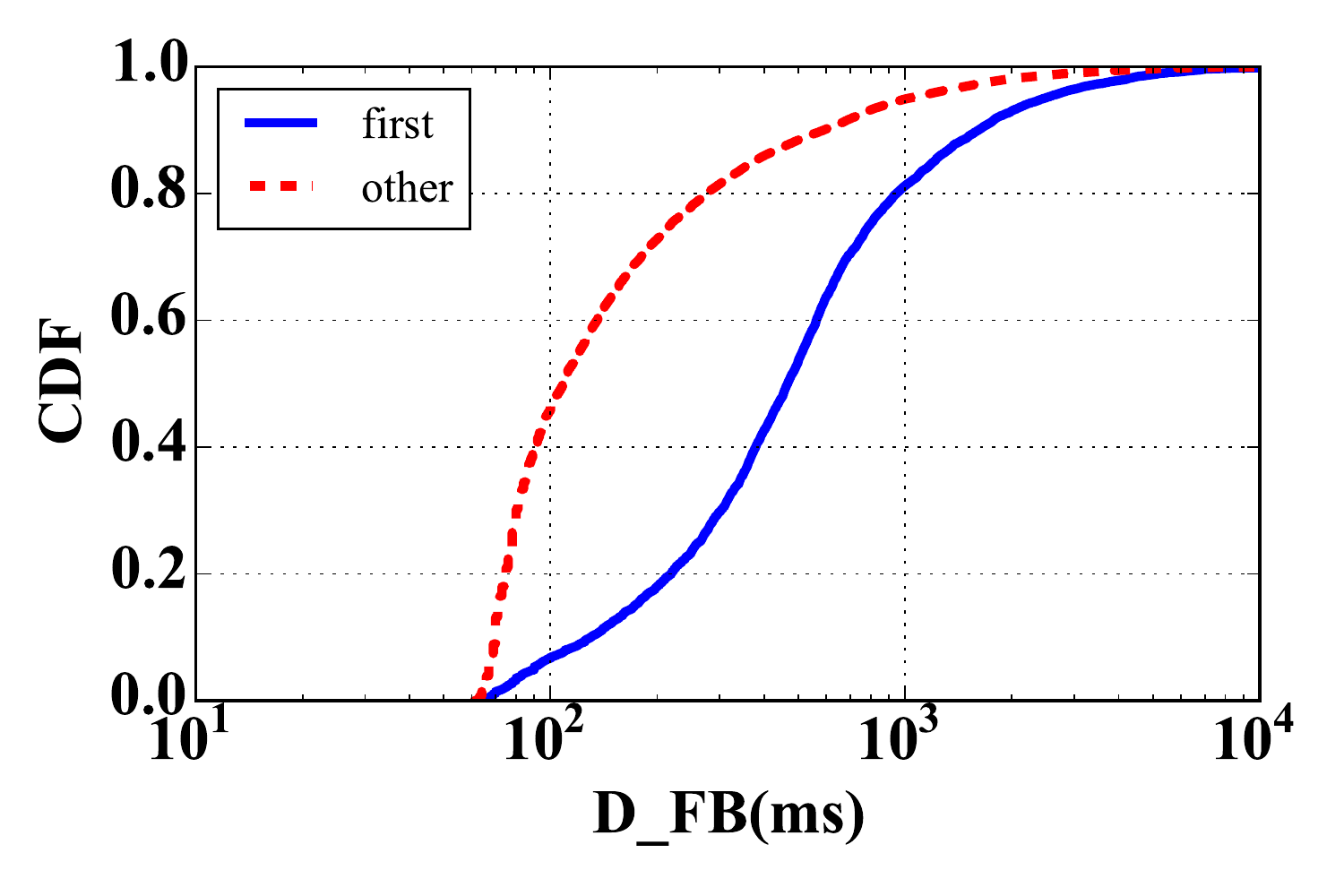}
\caption{$D_{FB}$ (ms) of first vs other chunks in equivalent performance conditions.}
\label{f:fb_chunkid}
\end{figure}

Figure~\ref{f:fb_chunkid} shows the CDF of $D_{FB}$ among the equivalent set  for first versus other chunks. We can see that despite similar performance conditions, first chunks still experience a higher $D_{FB}$. 
As the video data gets delivered to Flash, a progress event is fired to notify the player. Since an event listener has to be registered to listen to the progress events, the initialization of events and data path setup could increase $D_{FB}$ of first chunk. 
As an optimization, the polling frequency of the socket may be adjusted to adapt to the network arrival rate.~\footnote{We can only see Flash as a blackbox, hence, we cannot confirm this. However, a similar issue about ProgressEvent has been reported~\cite{flashdoc}.}
 
\textbf{Take-away:}
To make up for higher client-side latency of first chunks, we recommend the video providers to cache the first chunk of every video~\cite{752149} to eliminate other sources of performance problems and reduce the startup delay.

\subsection{Client's Rendering Stack}
\label{s:render}

\begin{figure}[!t]
\centering
  \includegraphics[width=0.8\linewidth]{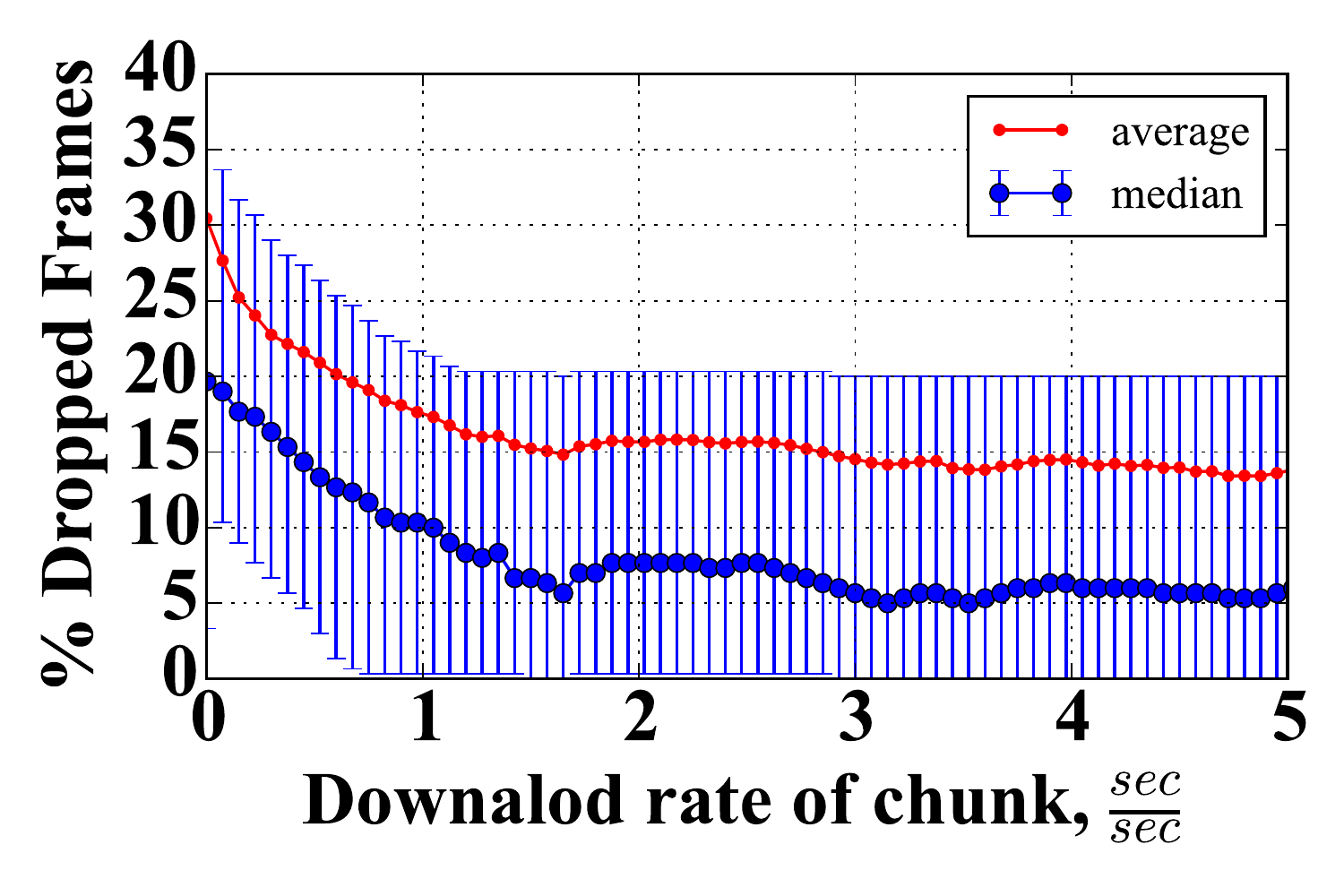}
  \caption{\%Dropped frames vs chunk download rate, first bar represents hardware rendering.}
  \label{f:avgfr_rate}
\end{figure}

\vspace{0.1in}\noindent\textbf{1. Avoiding dropped frames requires at least $1.5\frac{sec}{sec}$ download rate.}
The downloaded video chunks are a multiplexing of audio and video. They need to be de-multiplexed, decoded, and rendered on client's machine; which can take extra processing time on client's machine.  
When a chunk is downloaded fast, it provides more slack for the rendering path. We define the average download rate of a chunk as video length over download-time ($\frac{\tau}{D_{FB}+D_{LB}}$). 
Figure~\ref{f:avgfr_rate} shows the percent of dropped frames versus average download rate of chunks. A download rate of $1\frac{sec}{sec}$ is barely enough since after receiving the frames, more processing is needed to parse and decode frames for rendering. Increasing the download rate to $1.5\frac{sec}{sec}$ enhances the framerate; however, increasing the rate beyond this does not enhance the framerate more. 
To see if this observation can explain the rendering quality, we studied the framerate versus chunk download rate of chunks and observed that 85.5\% of chunks confirm the hypothesis, that is, they have bad framrate ($>30\%$ drop) when the download rate is below $1.5\frac{sec}{sec}$ and good framerate when download rate is at least $1.5\frac{sec}{sec}$. 5.7\% of chunks have low rates but good rendering, which can be explained by the buffered video frames that hide the effect of low rates. Finally, 6.9\% of chunks belong had low framerate despite a minimum download rate of $1.5\frac{sec}{sec}$, not confirming the hypothesis. However, this is explainable with following reasons:
First, \emph{the average download rate does not reflect instantaneous throughput}. In particular, the earlier chunks are more sensitive to fluctuations in throughput since fewer frames are buffered. Second, when \emph{the CPU on the client machine is overloaded}, software rendering can be inefficient \emph{regardless} of the chunk arrival rate. 

\begin{figure}[!t]
\centering
  \includegraphics[width=0.8\linewidth]{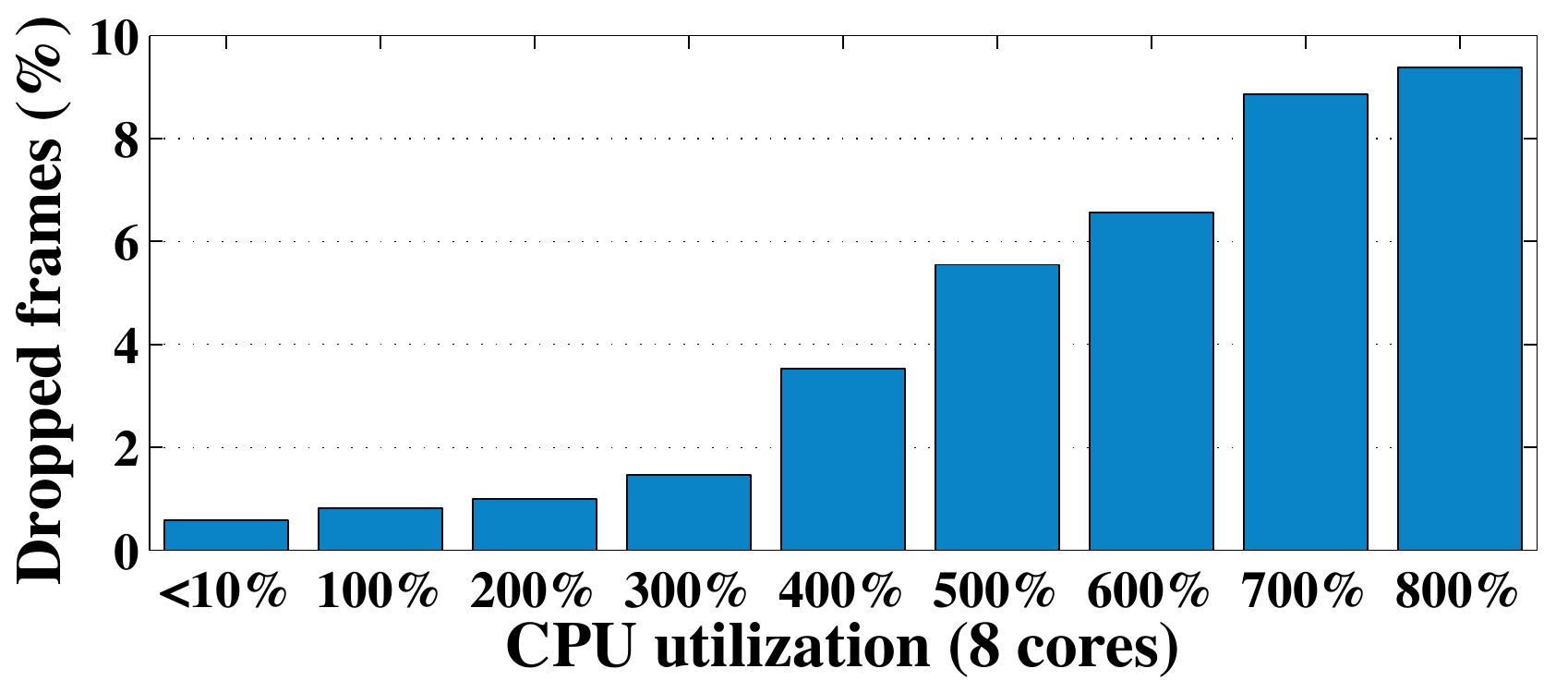}
  \caption{Dropped frames per CPU load in a controlled experiment.}
  \label{f:dropped_cpu}
\end{figure}

Figure~\ref{f:dropped_cpu} shows a simple controlled experiment: our player is running in Firefox browser on OS X with 8 CPU cores, connected to the server using a 1 GigE Ethernet, streaming a sample video with 10 chunks. The first bar represents the per-chunk dropped rate while using GPU. Next, we turned off hardware rendering to force rendering by CPU; at each iteration, we loaded one more CPU core.  

\vspace{0.1in}\noindent\textbf{2. Higher bitrates have better rendered framerate.}
Higher bitrates contain more data per frame, thus imposing a higher load on the CPU for decoding and rendering; thus, we expected chunks in higher bitrates to have more dropped frames as a result. We did not observe this expectation in our dataset. However, we observed the following trends in the data that can explain this paradox: (1) \emph{Higher bitrates are often requested in connections with lower RTT variation}: SRTTVAR across sessions with bitrates higher than 1Mbps is $5ms$ lower than the rest -- less fluctuations means less frames are delivered late. (2) \emph{higher bitrates are often requested in connections with lower retransmission rate}: the retransmission rate among sessions with bitrates higher than 1Mbps is more than $1\%$ lower than the rest -- lower packet loss rate means less frames are dropped or arrived late.

\vspace{0.1in}\noindent\textbf{3. Less popular browsers have worse rendering quality.}
If we limit our analysis to chunks with good performance ($rate > 1.5\frac{sec}{sec}$) that are also visible (i.e., $vis = True$), the rendering quality can still be bad due to inefficiencies in client's rendering path. Since we cannot measure the host load, we only characterize the clients based on their OS and browser.  

\begin{figure}[!t]
\centering
\includegraphics[width=80mm]{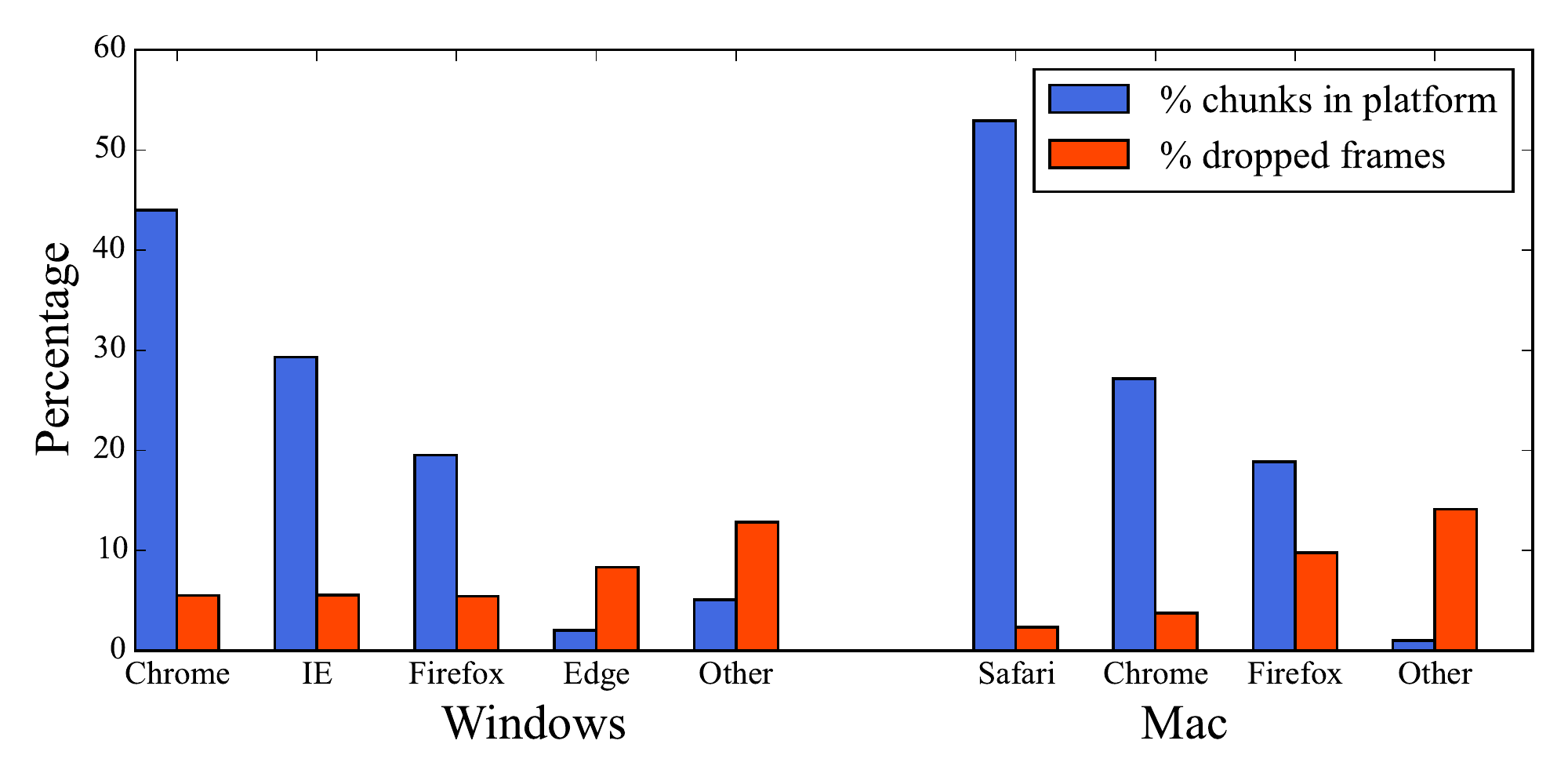}
\caption{Browser popularity and rendering quality in the two major platforms: Windows vs Mac.}
\label{f:browser_popularity}
\end{figure}

\begin{figure}[!t]
\centering
\includegraphics[width=80mm]{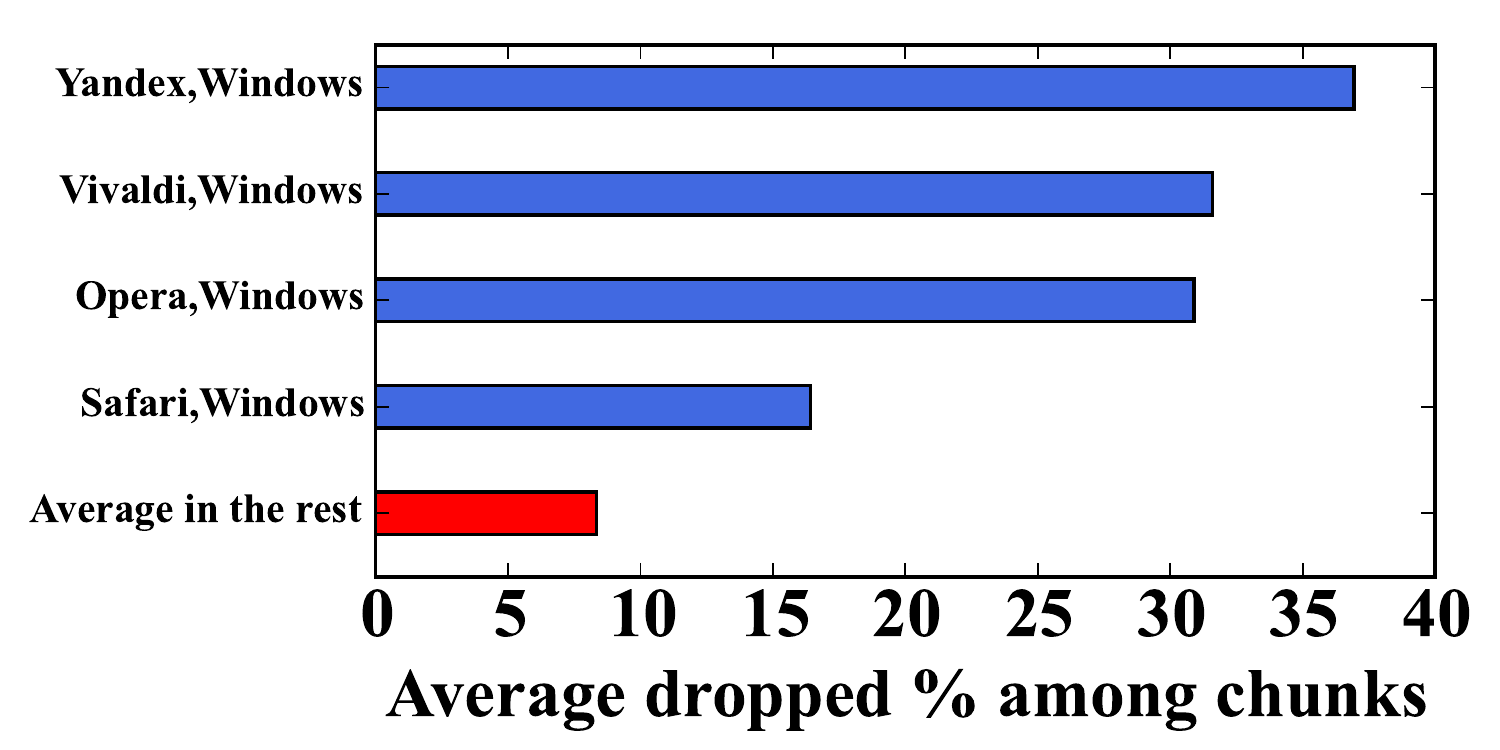}
\caption{Dropped \% of (browser, OS), $rate \geq 1.5\frac{sec}{sec}$, $vis = True$.}
\label{f:render_os_br}
\end{figure}

Figure~\ref{f:browser_popularity} shows the percentage of chunks requested from major browsers on each of OS X and Windows platforms (each platform is normalized to 100\%), as well as the average percentage of dropped frame of chunks among chunks served by that browser.  Among popular browsers, browsers with internal Flash (e.g., Chrome) and native HLS support (Safari on OS X) outperform other browsers (some of which may run Flash as a process, e.g., Firefox's protected mode). Also, the unpopular browsers (grouped in as Other) have the lowest performance, we further broke them down as shown in Figure~\ref{f:render_os_br}. These browsers have processed at least 500 chunks. Browsers such as Yandex, Vivaldi, Opera or Safari on Windows have the lowest rendered framerate compared to the average of  other browsers. 

\noindent\textbf{Take-aways:}
De-multiplexing, decoding, and rendering video chunks could be resource-heavy processes on client's machine. In absence of GPU (or hardware rendering), the burden falls on CPU to process frames efficiently; however, the resource demands from other applications can affect the rendering quality. We found that video rendering requires some processing time, and that a $1.5 \frac{sec}{sec}$ rate of video arrival can be used as a rule-of-thumb for achieving good rendering quality. Similar to download stack problems, rendering quality differs based on OS and browser. In particular, we found unpopular browsers to have lower rendering quality.

\section{Discussion}
Monitoring and diagnosis in large-scale content providers is a challenging problem due to insufficient instrumentation or measurement overhead limitations. In particular, (1) sub-chunk events such as bursty losses will not be captured in per-chunk measurements; and capturing them will impact player performance, (2) SRTT does not reflect the value of round-trip time at the time of measurement, rather is a smoothed average; to work within this limitation, we use methods discussed in (Section~\ref{s:net}); vanilla Linux kernels only export SRTTs to userspace today, 
(3) the characterization of the rendering path could improve by capturing the underlying resource utilization and environment, and (4) in-network measurements help further localization. For example, further characterization of network problems (e.g., is bandwidth limited at the core or the edge?) would have been possible using active probes (e.g., traceroute or ping) or in-network measurements from ISPs (e.g., link utilization). Some of these measurements may not be feasible at Web-scale.

\section{Related Work}

\vspace{0.1in}\noindent\textbf{Video streaming characterization:}
There is a rich area of related work in characterizing video-streaming quality. \cite{Plissonneau:2012:LVH:2155555.2155588} uses ISP packet traces to characterize video while~\cite{Yin:2009:IBN:1644893.1644946} uses CDN-side data to study content and Live vs VoD access patterns. Client-side data and a clustering approach is used in~\cite{Jiang:2013:SLS:2535372.2535394} to find critical problems related to user's ISP, CDN, or content provider.~\cite{Cha:2007:ITY:1298306.1298309} characterizes popularity in user-generated content video system. 
Our work differs from previous work by collecting and joining fine-grained per-chunk measurements from both sides and direct instrumentation of the video delivery path, including the client's download stack and rendering path.   

\vspace{0.1in}\noindent\textbf{QoE models:}
Studies such as~\cite{Dobrian:2011:UIV:2043164.2018478} have shown correlations between video quality metrics and user engagement.~\cite{Krishnan:2012:VSQ:2398776.2398799} shows the impact of video quality on user behavior using quasi experiments. Network data from commercial IPTV is used in ~\cite{Song:2011:QPS:2068816.2068836} to learn performance indicators for users QoE, where~\cite{Aggarwal:2014:PTQ:2565585.2565600} uses in-network measurements to estimate QoE for mobile users. 
We have used the prior work done on QoE models to extract QoE metrics that matter more to clients (e.g., the re-buffering and startup delay) to study the impact of performance problems on them.

\vspace{0.1in}\noindent\textbf{ABR algorithms:}
The bitrate adaptation algorithms have been studied well, ~\cite{Esteban:2012:IHA:2229087.2229094} studies the interactions between HTTP and TCP, while~\cite{Akhshabi:2011:EER:1943552.1943574} compares different algorithms in sustainability and adaptation. 
Different algorithms have been suggested to optimize video quality, in particular~\cite{Jiang:2012:IFE:2413176.2413189, Tian:2012:TAS:2413176.2413190} offer rate-based adaptation algorithms, where~\cite{Huang:2014:BAR:2619239.2626296} suggests a buffer-based approach, and~\cite{Yin:2015:CAD:2785956.2787486} aims to optimize quality using a hybrid model.
Our work is complementary to these works, because while an optimized ABR is necessary for good streaming quality, we showed problems where a good ABR algorithm is not enough and corrective actions from the content provider are needed.

\vspace{0.1in}\noindent\textbf{Optimizing video quality by CDN selection:}
Previous work suggests different methods for CDN selection to optimize video quality, for example~\cite{Torres:2011:DVS:2014697.2014851} studies policies and methods used for server selection in Youtube, while~\cite{Krishnan:2009:MBE:1644893.1644917} studies causes of inflated latency for better CDN placement. 
Some studies~\cite{Liu:2012:CCI:2342356.2342431, Georgopoulos:2013:TNQ:2491172.2491181, Ganjam:2015:CIC:2789770.2789780} make the case for centralized video control planes to dynamically optimize the video delivery based on a global view while~\cite{Balachandran:2013:APB:2504730.2504743} makes the case for federated and P2P CDNs based on content, regional, and temporal shift in user behavior.

\section{Conclusion}
In this paper, we presented the first Web-scale and end-to-end measurement study of the video streaming path to characterize problems located at a content provider's CDN, network, and the client's download and rendering paths.  Instrumenting the end-to-end path gives us a unique opportunity to look at multiple components together during a session, at per-chunk granularity, and to discover transient and persistent problems that affect the video streaming experience. We characterize several important characteristics of video streaming services, including causes for persistent problems at CDN servers such as unpopularity, sources of persistent high network latency and persistent rendering problems caused by browsers. We showed that we can draw insights into the client's download stack latency (possible at scale only via e2e instrumentation); and the download stack can impact the QoE and feed incorrect information into the ABR algorithm. We discussed the implications of our findings for content providers (e.g., pre-fetching subsequent chunks), ISPs (establishing better peering points), and the ABR logic (e.g., using the apriori observations about client prefixes).

\section*{Acknowledgments}
This work benefited from a large number of folks from the Video Platforms Engineering group at Yahoo. We thank the group for helping us with instrumentation and patiently answering many of our questions (and hearing our performance optimization observations). Thanks to P.P.S. Narayan for supporting this project at Yahoo.

{\footnotesize \bibliographystyle{acm}
\bibliography{ref}}
\end{document}